\documentclass[twocolumn]{aastex63}
\usepackage{multirow}
\usepackage{amsmath}
\usepackage{booktabs}
\usepackage{wrapfig}

\usepackage{xcolor}

\newcommand\lsim{\mathrel{\rlap{\lower4pt\hbox{\hskip1pt$\sim$}}
\raise1pt\hbox{$<$}}}
\accepted{ApJ}
\shorttitle{SIGOs' evolutionary tracks}
\shortauthors{Lake et al.}
\graphicspath{{./}{figures/}}

\begin{document}

\title{ The Supersonic Project: The Early Evolutionary Path of SIGOs }


\correspondingauthor{William Lake}
\email{wlake@astro.ucla.edu}
\author[0000-0002-4227-7919]{William Lake}
\affil{Department of Physics and Astronomy, UCLA, Los Angeles, CA 90095}
\affil{Mani L. Bhaumik Institute for Theoretical Physics, Department of Physics and Astronomy, UCLA, Los Angeles, CA 90095, USA\\}

\author[0000-0002-9802-9279]{Smadar Naoz}
\affil{Department of Physics and Astronomy, UCLA, Los Angeles, CA 90095}
\affil{Mani L. Bhaumik Institute for Theoretical Physics, Department of Physics and Astronomy, UCLA, Los Angeles, CA 90095, USA\\}

\author[0000-0001-5817-5944]{Blakesley Burkhart}
\affiliation{Department of Physics and Astronomy, Rutgers, The State University of New Jersey, 136 Frelinghuysen Rd, Piscataway, NJ 08854, USA \\}
\affiliation{Center for Computational Astrophysics, Flatiron Institute, 162 Fifth Avenue, New York, NY 10010, USA \\}

\author[0000-0003-3816-7028]{Federico Marinacci}
\affiliation{Department of Physics \& Astronomy ``Augusto Righi", University of Bologna, via Gobetti 93/2, 40129 Bologna, Italy\\}

\author[0000-0001-8593-7692]{Mark Vogelsberger}
\affil{Department of Physics and Kavli Institute for Astrophysics and Space Research, Massachusetts Institute of Technology, Cambridge, MA 02139, USA\\}

\author[0000-0001-6246-2866]{Gen Chiaki}
\affiliation{Astronomical Institute, Tohoku University, 6-3, Aramaki, Aoba-ku, Sendai, Miyagi 980-8578, Japan}

\author[0000-0003-4962-5768]{Yeou S. Chiou}
\affil{Department of Physics and Astronomy, UCLA, Los Angeles, CA 90095}
\affil{Mani L. Bhaumik Institute for Theoretical Physics, Department of Physics and Astronomy, UCLA, Los Angeles, CA 90095, USA\\}

\author[0000-0001-7925-238X]{Naoki Yoshida}
\affiliation{Kavli Institute for the Physics and Mathematics of the Universe (WPI), UT Institute for Advanced Study, The University of Tokyo, Kashiwa, Chiba 277-8583, Japan}
\affiliation{Department of Physics, The University of Tokyo, 7-3-1 Hongo, Bunkyo, Tokyo 113-0033, Japan}

\author[0000-0002-0984-7713]{Yurina Nakazato}
\affiliation{Department of Physics, The University of Tokyo, 7-3-1 Hongo, Bunkyo, Tokyo 113-0033, Japan}

\author[0000-0003-2369-2911]{Claire E. Williams}
\affil{Department of Physics and Astronomy, UCLA, Los Angeles, CA 90095}
\affil{Mani L. Bhaumik Institute for Theoretical Physics, Department of Physics and Astronomy, UCLA, Los Angeles, CA 90095, USA\\}



\begin{abstract}
Supersonically Induced Gas Objects (SIGOs) are a class of early Universe objects that have gained attention as a potential formation route for globular clusters. SIGOs have recently begun to be studied in the context of molecular hydrogen cooling, which is key to characterizing their structure and evolution. Studying the population-level properties of SIGOs with molecular cooling is important for understanding their potential for collapse and star formation, and for addressing whether SIGOs can survive to the present epoch. Here, we investigate the evolution of SIGOs before they form stars, using a combination of numerical and analytical analysis. We study timescales important to the evolution of SIGOs at a population level in the presence of molecular cooling. Revising the previous formulation for the critical density of collapse for SIGOs allows us to show that their prolateness tends to act as an inhibiting factor to collapse. We find that simulated SIGOs are limited by artificial two-body relaxation effects that tend to disperse them. We expect that SIGOs in nature will be longer-lived compared to our simulations. Further, the fall-back timescale on which SIGOs fall into nearby dark matter halos, potentially producing a globular-cluster-like system, is frequently longer than their cooling timescale and the collapse timescale on which they shrink through gravity. Therefore, some SIGOs have time to cool and collapse outside of halos despite initially failing to exceed the critical density. From this analysis we conclude that SIGOs should form stars outside of halos in non-negligible stream velocity patches in the Universe.
\end{abstract}

\keywords{Globular star clusters --- High-redshift galaxies --- Giant molecular clouds --- Star formation --- Galactic and extragalactic astronomy}

\section{Introduction}\label{sec:intro}

The standard model of structure formation, invoking dark energy and dark matter (DM) that dominate the Universe's energy density budget, has had great success in explaining a wide variety of observations. The $\Lambda$CDM model successfully explains the anisotropies in the Universe's radiation field and the large-scale distribution of galaxies, as well as more generally explaining properties of the Universe's structure on scales larger than $100$ Mpc \citep[][]{Springel+05,Vogelsberger+14a,Vogelsberger+14b,Schaye+15}. This model predicts that the Universe's structure formed hierarchically, with very early primordial baryon overdensities at $z \lessapprox 30$ collapsing to form larger objects, which eventually form galaxies and other structures. 

These early baryon overdensities evolved in turn from interactions with existing DM overdensities following recombination. Because baryon objects had been inhibited from collapse prior to recombination by the photon field, DM overdensities at the time of recombination were about $10^5$ times larger than baryon overdensities \citep[e.g.,][]{Naoz+05}. Because of this imbalance, the formation and growth of baryon overdensities at this time was driven by these large existing DM overdensities. 

However, this process was complicated by the fact that there was a significant relative velocity between baryons and DM following recombination \citep{TH}. This relative velocity had an rms value at recombination of $30$ km s$^{-1}$ (which was about $5$ times the average speed of sound in the Universe at this time) and was coherent on few-Mpc scales. Because of its coherence on these scales, it is also known as the stream velocity.

Because the stream velocity was so highly supersonic, it induced significant spatial offsets between baryon overdensities and their parent DM overdensities, forming collapsed baryon objects outside the virial radii of their parent DM halos \citep{naoznarayan14}. These collapsed baryon objects, known as Supersonically Induced Gas Objects or SIGOs, have been proposed as a progenitor of some modern-day globular clusters (GCs), and have been shown to have similar properties to GCs \citep[e.g.,][]{chiou18,chiou+19,Chiou+21,Lake+21,Nakazoto+22}. However, SIGOs remain poorly understood. Early simulations of SIGOs, aimed at confirming their existence and putting basic constraints on their properties \citep[e.g.,][]{popa,chiou18,chiou+19,Chiou+21,Lake+21}, included only adiabatic and sometimes atomic hydrogen cooling. However, molecular hydrogen cooling has important effects on the structure of SIGOs \citep{Nakazoto+22}, and on the abundance of gas objects in general in the early Universe \citep[e.g.][]{Glover13,Schauer+21,Nakazoto+22,Conaboy+22}. For example, SIGOs in molecular cooling simulations tend to be far more filamentary than those in atomic cooling simulations \citep{Nakazoto+22}, potentially for reasons analogous to the Zel'dovich approximation, which states that collapse will occur along successive axes, starting with the shortest \citep{Zeldovich}. H$_2$ cooling also lowers the temperature in these SIGOs to $\sim200$ K, which lowers the Jeans mass to about $10^3$ M$_\odot$, potentially leading to star formation in SIGOs \citep[][Lake et al. in prep]{Nakazoto+22}. 

In order to understand and contextualize the properties of SIGOs with molecular hydrogen cooling, it is useful to have an analogous class of objects to compare to, which share many of the properties of SIGOs. A natural class of objects to compare to here are giant molecular clouds (GMCs). Like SIGOs, GMCs form in environments shaped by supersonic turbulence \citep[e.g.,][]{Larsen+81, Padoan+99,MacLow+04,Krumholz+05, McKee+07, Krumholz+07,Bergin+07,Padoan+11,Burkhart+12,Semenov+16,Mocz+17,burkhart18a,Burkhart+21,Appel+22}. The energy from this turbulence cascades from the driving scale of the turbulence ${\rm L}_{\rm Drive}$ down to the sonic scale $\lambda_s$  which marks the boundary between supersonic and subsonic turbulence: 
\begin{equation} 
\lambda_s = \frac{\rm{L}_{\rm Drive}}{\mathcal{M}^2} \ ,
\end{equation} 
where $\mathcal{M}$ is the mach number of the turbulent flow on the driving scale, and where we have assumed Larson's law \citep{Larsen+81}. This energy cascade is important in both cases, because it creates high density peaks in the turbulent medium on scales comparable to the sonic scale \citep[e.g.,][]{Krumholz+05}, which is key to understanding both objects' structure. Ultimately, this process leads to a critical density for star formation that can be expressed in the form 
\begin{equation}\label{Eq:rhocrit}
\rho_{\rm crit, GMC} = \frac{\pi c_s^2 \mathcal{M}^4} {G {\rm L}_{\rm Drive}^2} \ ,
\end{equation}
where $c_s$ is the speed of sound and $G$ is the gravitational constant.

GMCs are also interesting as a point of comparison for SIGOs because they may be a major formation channel for globular clusters at early times \citep[e.g., $z\sim 6$][]{Harris+94,Elmegreen+97,Kravtsov+05, Shapiro2010, Grudic+22}. GMCs form with a wide range of masses and densities, and it has been shown that the high-density, high-mass end of this formation channel may be capable of efficiently forming globular clusters. This formation process is supported by observations of the merging Antennae system and its population of massive young clusters \citep[e.g.,][]{Whitmore+95,Whitmore+99}. Therefore, a better understanding of the comparison between SIGOs and GMCs may also be important to understanding the link between SIGOs and globular clusters.

Motivated by these factors, in the present paper we present a suite of simulations using the cosmological simulation code {\tt AREPO}, including primordial chemistry and accounting for the effects of molecular hydrogen cooling. We constrain SIGOs' structural and kinematic properties, including the effects of the stream velocity and supersonic turbulence on the size and density of SIGOs, and compare these properties to those of GMCs to better contextualize them. As the formation of SIGOs is now well-understood, we aim to characterize the next step of SIGOs' lives by providing physical intuitions for the processes through which SIGOs evolve. To this end, we provide an analysis of the various timescales important for early SIGO evolution. This includes the cooling time, as well as the timescale on which they collapse gravitationally. It also includes the free-fall time on which SIGOs fall into nearby DM halos. Taken together, these characteristic timescales provide a better understanding of the potential for star formation in SIGOs. We also discuss two other mechanisms relevant to SIGO evolution: growth and two-body relaxation.

This paper is organized as follows: in Section~\ref{sec:Meth} we discuss the setups of the simulations used, as well as modifications made to definitions of SIGOs used in prior works. In Section~\ref{Sec:MolecCool} we expand upon the importance of molecular hydrogen cooling to SIGOs' properties. In Section~\ref{Sec:Results}, we discuss the similarities and differences between SIGOs and GMCs, and revisit previous density-driven analyses of SIGO evolution. In Section~\ref{sec:evolution}, we overview the evolution of SIGOs from a timescale perspective, primarily discussing the potential for SIGOs to collapse through the lens of molecular hydrogen cooling. Lastly, in Section~\ref{Sec:Discussion} we overview future avenues of exploration towards understanding star formation in SIGOs and summarize our results.

For this work, we have assumed a $\Lambda$CDM cosmology with $\Omega_{\rm \Lambda} = 0.73$, $\Omega_{\rm M} = 0.27$, $\Omega_{\rm B} = 0.044$, $\sigma_8  = 1.7$, and $h = 0.71$.

\section{Numerical Setup and Object Classification}\label{sec:Meth}
Using the cosmological and hydrodynamic simulation code {\tt AREPO}, we present a suite of $4$ simulations demonstrating the effect of molecular hydrogen cooling on the properties of SIGOs. The main parameters of these simulations are listed in Table~\ref{Table:Sims}. Here, we indicate Runs with molecular hydrogen cooling using ``H$2$", and indicate runs without molecular hydrogen cooling (simply using adiabatic and atomic hydrogen cooling) with ``H". Runs without the stream velocity are labelled with ``$0$v", whereas Runs with a $2\sigma$ stream velocity (a 2 $v_{\rm rms}$ relative velocity between baryons and dark matter) are labelled as ``$2$v". The stream velocity in the latter Runs is implemented as a uniform boost to baryon velocities in the x direction in the initial conditions--at the initial redshift of $z=200$, this is a boost of $11.8 $ km s$^{-1}$. 

Our initial conditions were generated using transfer functions calculated using a modified version of {\tt CMBFAST} \citep{seljak96} taking into account the first-order correction of scale dependent temperature fluctuations \citep[e.g.,][]{NB,naoznarayan13} and second-order corrections to the equations presented in \citet{TH}, describing the evolution of the stream velocity. We use two transfer functions, one each for the baryons and DM, as the evolution of the gas fraction strongly depends on the initial conditions of the baryons \citep[e.g.,][]{Naoz+09, naoz11,naoz12, Naoz+13, park+20}. The glass file for baryons uses positions shifted by a random vector, rather than tracing the dark matter density perturbations \citep{Yoshida+03b}.

Our simulations use $512^3$ DM particles with a mass of $1.9\times 10^3$ M$_\odot$ and $512^3$ Voronoi mesh cells corresponding to a gas mass of $360$ M$_\odot$, in a box $2$ comoving Mpc on a side. This box size aims to study a patch of the Universe with constant stream velocity, as a proof of concept, rather than analysing structure formation in larger regions with variable stream velocity. Following \citet{chiou+19,Chiou+21, Lake+21, Nakazoto+22}, we choose $\sigma_8 = 1.7$, representing a rare, overdense region where structure forms early in a large volume, in order to increase the number of gas objects in our simulation, allowing increased statistical power. The simulations begin at $z=200$, and they run to $z=20$. 

In Runs $0$vH$2$ and $2$vH$2$, using the chemistry and cooling library GRACKLE \citep{Smith+17, Chiaki+19}, we track nonequilibrium chemical reactions and their associated radiative cooling explicitly in the gas. This Run includes H$_2$ and HD molecular cooling, as well as chemistry for 15 primordial species: e$^-$, H, H$^+$, He, He$^+$, He$^{++}$, H$^-$, H$_2$, H$_2^+$, D, D$^+$, HD, HeH$^+$, D$^-$, and HD$^+$. The radiative cooling rate of H$_2$ follows both rotational and vibrational transitions \citep{Chiaki+19}. 
\begin{figure*}[t]

\centering

\gridline{\fig{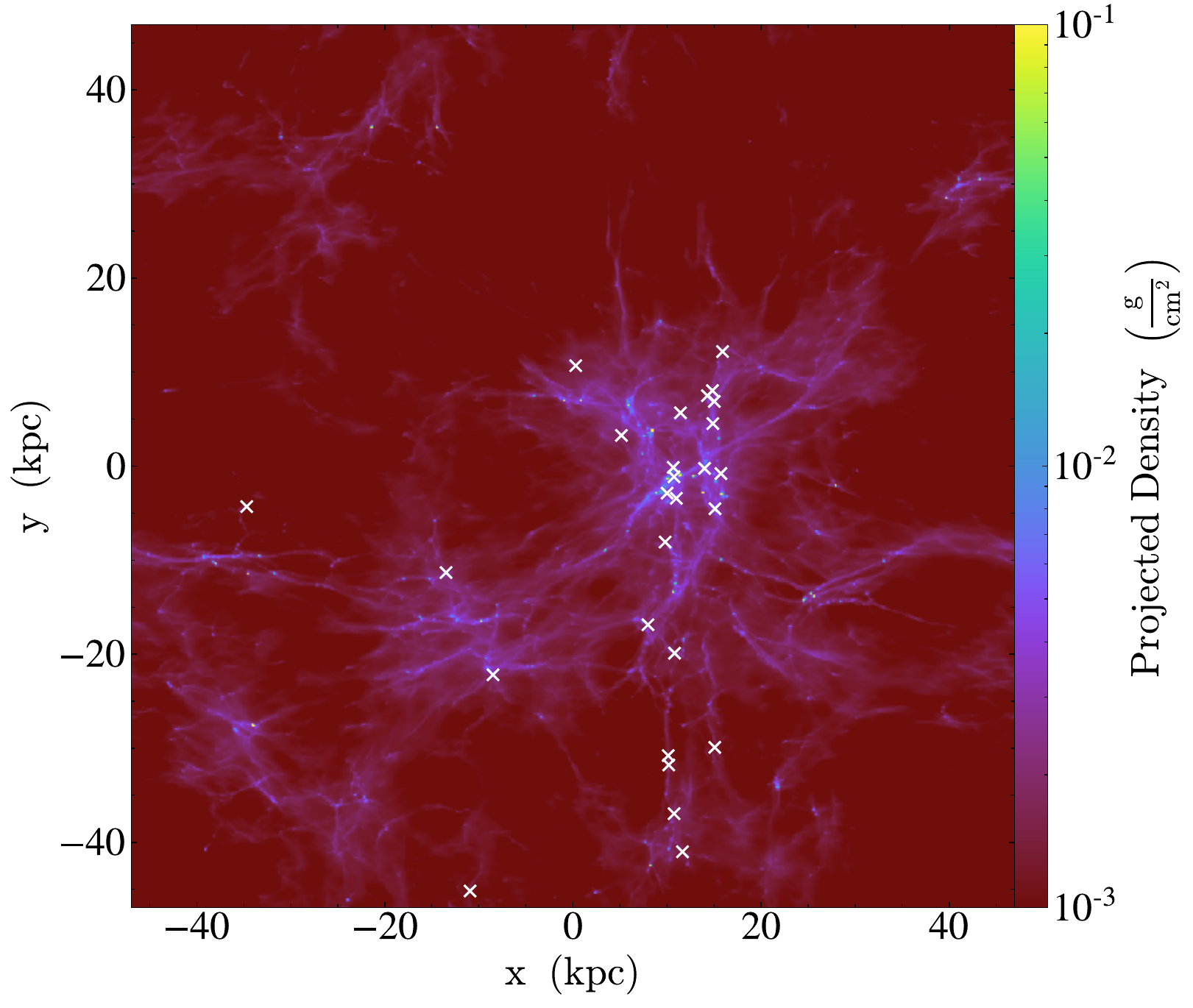}{0.47\textwidth}{\centering Atomic Cooling Only}
          \fig{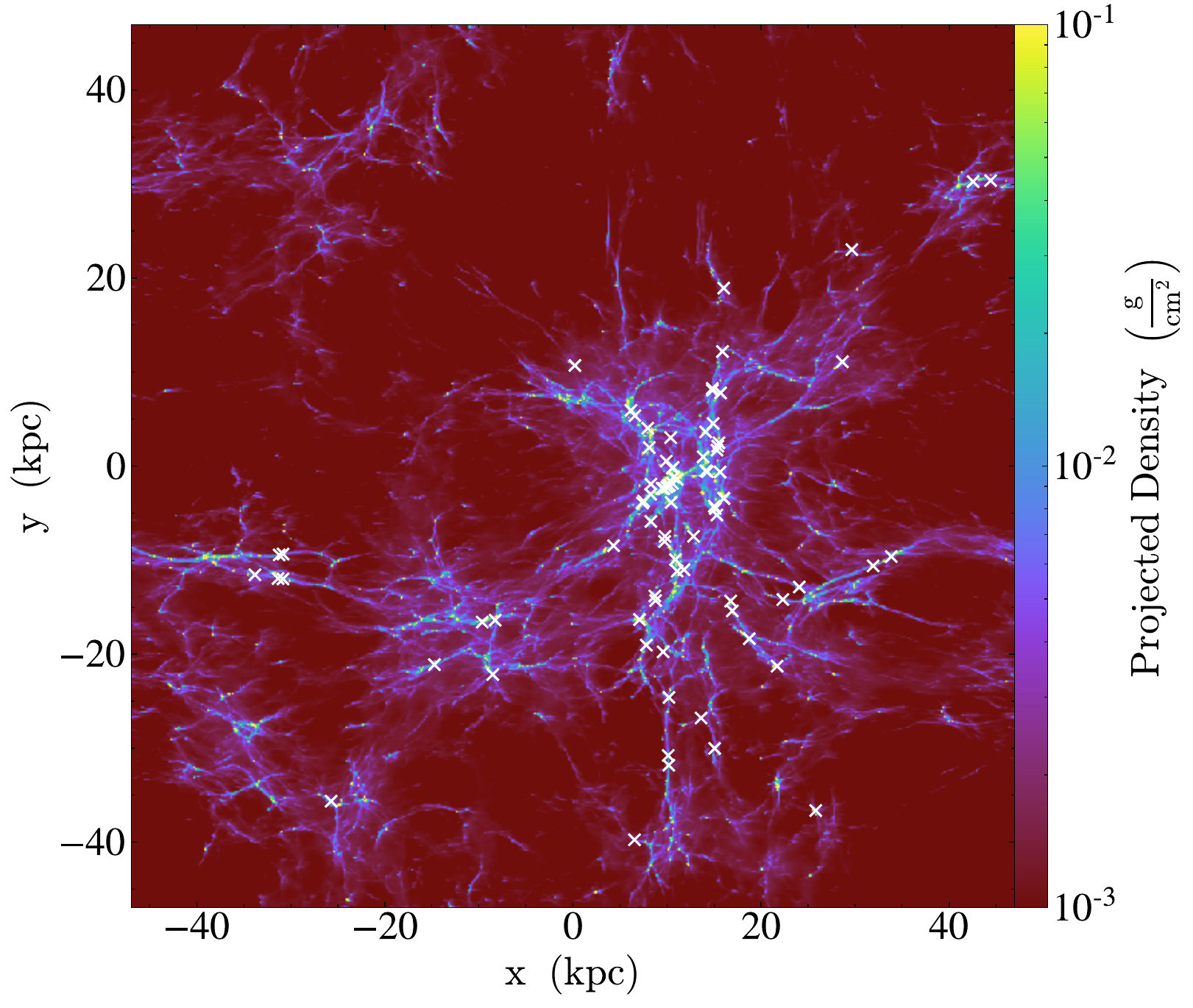}{0.47\textwidth}{\centering Molecular Hydrogen Cooling}
          }
    \caption{{\bf Comparison of the abundance of SIGOs with and without molecular cooling at $z=20$}. We show the gas density field in our simulation box for simulation $2$vH$2$ (molecular cooling, right panel), compared to simulation $2$vH (atomic cooling, left panel). SIGOs are marked with Xs. Note here that SIGOs trace gas and halo abundances on these scales. Note also that molecular hydrogen cooling dramatically increases the abundance of SIGOs. }%
    \label{fig:FullMapSig2}%
\end{figure*}
For comparison, in Runs $0$vH and $2$vH, we consider only atomic hydrogen cooling, following the species e$^-$, H, H$^+$, He, He$^+$, and He$^{++}$ in equilibrium with a spatially uniform, redshift-dependent photoionizing background, as described in \citet{Vogelsberger+13}.

We use the object classifications laid out in \citet{chiou18}. We identify DM halos with an FOF (friends-of-friends) algorithm that uses a linking length of $20\%$ of the mean particle separation of the DM component, which is about $780$ comoving pc. Assuming sphericity for simplicity, we use this to calculate positions and radii of all DM halos in the simulation output \citep[though note that DM halos at this time resemble triaxial ellipsoids, e.g.][]{Sheth+01, Lithwick+11, Vogelsberger+11, Schneider+12, Vogelsberger+20}.

We then run the FOF finder on the gas component, allowing us to identify gas-primary objects that contain over 100 gas cells. Especially when molecular hydrogen is included, SIGOs at these redshifts are distinctly filamentary, similar to in the Zel'dovich approximation \citep{Zeldovich}. We cannot, therefore, assume sphericity when determining these objects' properties, as we did with DM halos. To address this issue, we follow the procedure introduced in \citet{popa}. In particular, we fit these gas-primary objects to ellipsoids determined as the smallest ellipsoidal surface that surrounds all of the constituent gas cells. We then reduce the major axis of this ellipsoid by $5$\% until either the ratio of the axes lengths of the tightened ellipsoid to that of the original ellipsoid is greater than the fraction of gas cells remaining inside the tightened ellipsoid, or until 20\% of their particles have been removed. Finally, in order to distinguish SIGOs from other classes of gas objects, we require that the center of mass of the gas objects be located outside the virial radius of all nearby halos, and that the SIGOs have a baryon fraction above $60$\%.\footnote{This condition was introduced as $40$\% in initial work by \citet{chiou18}, and modified by \citet{Nakazoto+22} to account for false identifications of SIGOs arising from molecular cooling, as discussed in Section~\ref{Sec:MolecCool} and justified in Appendix~\ref{Sec:appendix}.} This condition is discussed and justified in Appendix~\ref{Sec:appendix}.

\begin{table}[htbp]
\centering
\begin{tabular}{ccc}
Run & $v_\mathrm{bc}$ & H$_2$ Cooling \\\hline
0vH2 & 0            & Yes        \\\hline
0vH  & 0            & No         \\\hline
2vH2 & 2$\sigma$    & Yes        \\\hline
2vH  & 2$\sigma$    & No         \\\hline

\end{tabular}
 \caption{Simulation Parameters}
  \label{Table:Sims}
\end{table}

After identifying every SIGO present in each of the 150 snapshots in Run $2$vH$2$ (evenly spaced in redshift from $z=30$ to $z=20$), we compare the gas cells present in each gas-primary object at each redshift to track gas objects across snapshots. If two gas objects in different snapshots share at least $1/3$ of the gas cells present in the smaller of the two objects, they are assumed to be the same object at different times. In other words, we track the gas cells’ IDs and require that at least $1/3$ of the gas cells will be shared by the larger gas object. By identifying which of these objects are SIGOs (and at what redshifts they fulfill the conditions needed to be identified as a SIGO), we can trace the history of SIGOs in Run $2$vH$2$.

\section{The Importance of Molecular Cooling to SIGOs' Properties}\label{Sec:MolecCool}

Molecular hydrogen cooling plays a vital role in the formation of the first stars \citep[e.g.,][]{Saslaw+67,Haiman+96,Abel+98,Yoshida+03early,Stacy+10,greif11,Yoshida+08,Bromm13,Glover13,Hummel+16,Nakazoto+22}. Previous studies of SIGOs have argued that adiabatic and atomic cooling may be a sufficient approximation to understand SIGOs' formation and morphology \citep[e.g.,][]{popa,chiou18,Chiou+21,Lake+21}. This is because the main factor in the formation of SIGOs is the phase shift of  gravitational fluctuations between DM and baryons \citep{naoznarayan14}. However, it was recently shown that molecular cooling may be able to increase the efficiency of SIGO formation, as well as playing an important role in their density \citep{Schauer+21,Nakazoto+22}.

Quantifying the population-level differences in SIGOs' properties with molecular cooling compared to atomic cooling can help to illuminate the physical processes important to their later evolution. Figure~\ref{fig:FullMapSig2} shows a comparison of SIGO abundances with and without molecular hydrogen cooling (Runs $2$vH and $2$vH$2$). SIGOs in this figure are marked with white Xs, displayed against a backdrop of the gas density in the full simulation box. Note that length scales here are in physical kpc. With molecular hydrogen cooling, SIGOs are dramatically more efficient at condensing from overdensities in the gas. Thus, we find $85$ SIGOs at $z=20$ in Run $2$vH$2$, compared to $27$ in Run $2$vH. With molecular cooling, SIGOs also reach higher densities, due to their increased cooling efficiencies. In Run $2$vH$2$, the most dense SIGO at $z=20$ had an overall gas density of $39.1$ cm$^{-3}$, compared to a maximum gas density in any SIGO of $2.5$ cm$^{-3}$ in Run $2$vH. Therefore, SIGOs are more abundant and denser by redshift $20$ with molecular cooling.

\begin{figure}[t]
\includegraphics[width=1\linewidth]{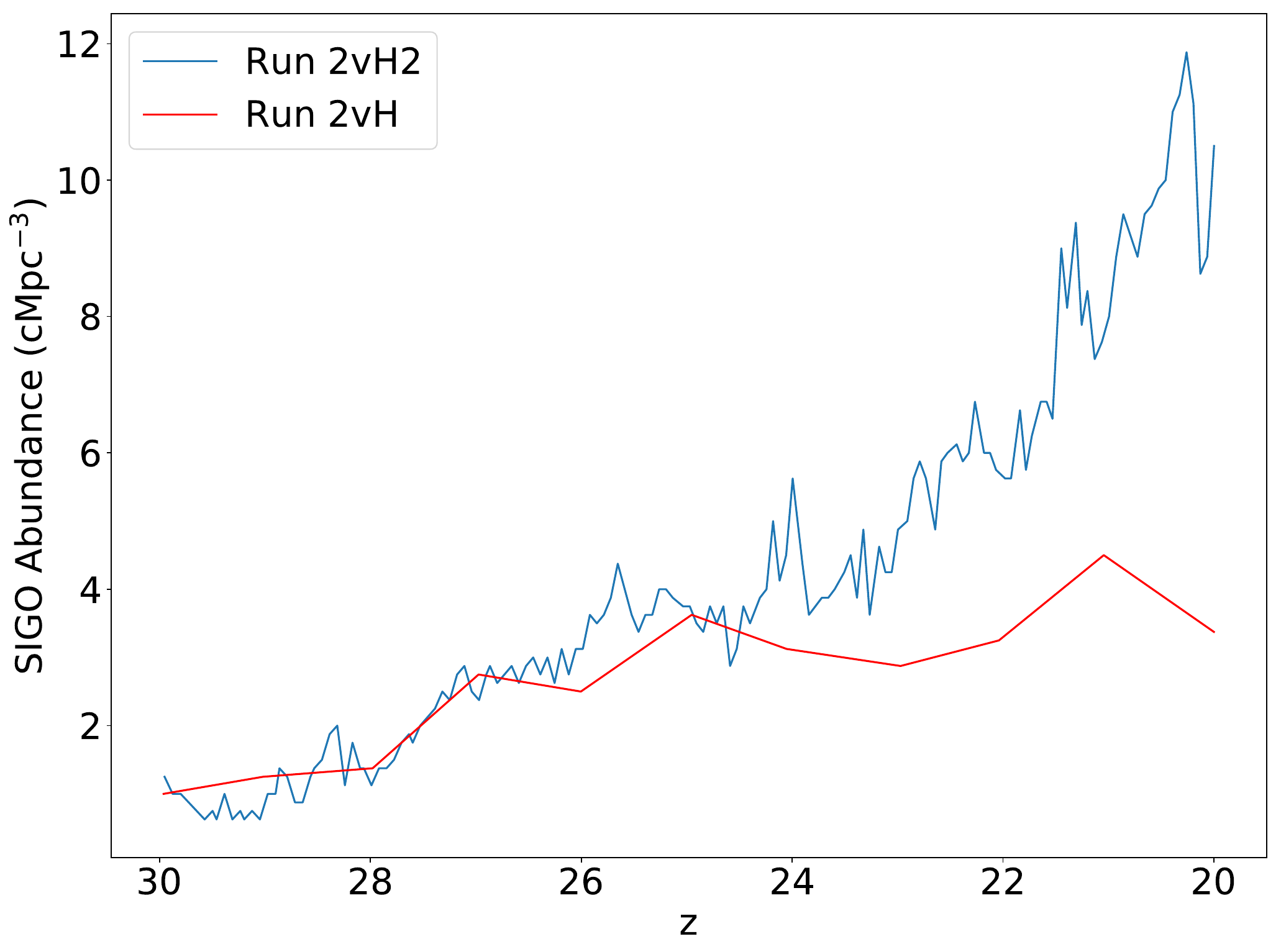}
\caption{{\bf Changes in SIGO abundance with redshift.} Plotted here is the evolution of SIGO abundances with redshift in Run $2$vH$2$. For comparison, SIGO abundances from Run $2$vH are plotted in red. Note that the time resolution of Run $2$vH$2$ is higher than that of Run $2$vH, for the purposes of this study.  }\label{fig:SIGOAbundances}
\end{figure}

The increase in SIGO abundances through molecular cooling seen in Figure~\ref{fig:FullMapSig2} can be seen as a function of redshift in Figure~\ref{fig:SIGOAbundances}. This figure shows the time evolution of SIGO abundances with molecular cooling, as well as showing SIGO abundances at integer redshifts in Run $2$vH for comparison. As one can see, the abundance of SIGOs increases with time, and consistent with \citet{Nakazoto+22}, abundances are generally enhanced through molecular cooling, through the process outlined above. This process is most important at later redshifts: as one can see in the figure, the abundance of SIGOs in Run $2$vH$2$ begins to significantly diverge from the abundance without molecular cooling after $z \sim 25$ in our simulations.

Molecular cooling affects the structure around SIGOs as well: SIGOs form embedded in gas filaments, and molecular hydrogen cooling permits these filaments to condense much more efficiently than does atomic hydrogen cooling. As mentioned in Section~\ref{sec:Meth}, following \citet{Nakazoto+22}, we revise the gas fraction cutoff for identification of SIGOs with our FOF algorithm upwards compared to e.g. \citet{Chiou+21}, as we are working with molecular hydrogen cooling. This exclusively reduces the number of SIGOs considered, in order to prioritize positive identifications over concerns about false negatives, allowing us to state with confidence that objects studied are truly SIGOs. This is because gas filaments condense so efficiently that they sometimes appear to structure-finding algorithms (even in Run $0$vH$2$ where no SIGOs should form) as collapsed baryon objects outside of halos. Because SIGOs are embedded in these gas filaments, this process is inextricably linked to their formation: the mass and particularly the density of cooled gas around SIGOs in these filaments increases in Run $2$vH$2$ compared to Run $2$vH, potentially allowing SIGOs to accrete external gas and grow further with molecular cooling. 

\begin{figure*}[t]

\centering

\gridline{\fig{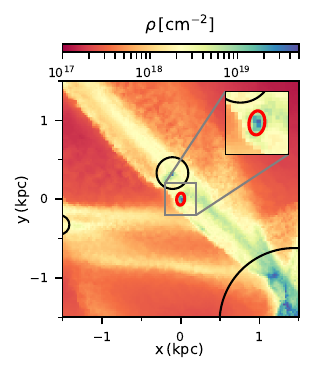}{0.47\textwidth}{\centering Gas Column Number Density}
          \fig{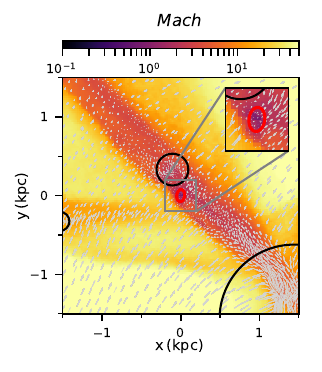}{0.47\textwidth}{\centering Mach number}
          }
\caption{\textbf{Example of a Young SIGO in a Turbulent Shock:} The velocity field of a typical SIGO (in red) with molecular cooling. This SIGO is associated with a nearby parent DM halo, and there is a larger DM halo at the bottom right of the field. Here, we see that the velocity dispersion inside the SIGO is quite small compared to that outside of it. The sonic scale here is larger than the scale of the SIGO, so turbulent flow plays a small role in the SIGO's potential further collapse. As can be seen the SIGO is embedded in a shock front, which has formed on a scale where the Mach number is about unity. See Figure~\ref{fig:VelocitySphericalAverage}.  }\label{fig:ExampleSIGOMach}
\end{figure*}

\section{SIGOs as GMC analogues}\label{Sec:Results}

As mentioned above (\S \ref{sec:intro}), at face value, GMCs and SIGOs seem to share several key properties. At the most fundamental level, the impact of molecular hydrogen is critical for understanding the evolution of both classes of objects, which also have similar masses ($\sim 10^6$ M$_\odot$) and scales ($\sim 100$ pc). Additionally, these classes have similar substructures, with SIGOs often exhibiting local density fluctuations such as cores and filaments in simulations, similar to those in GMCs \citep{MacLow+04,Krumholz+05,Mocz+18}. In addition, the formation of both classes of object is dominated by supersonic turbulence from their environment \citep{Burkhart+09,Padoan+11,Burkhart+12,Semenov+16,Mocz+17,burkhart18a,chiou+19,Burkhart+21,Appel+22}. In considering star formation within both classes of objects, it is often impossible to consider only the Jeans mass-- instead, one must use a critical density that incorporates the impact of turbulence. 

To illustrate the nature of the turbulence around SIGOs, in Figure~\ref{fig:ExampleSIGOMach}, we see the velocity field around one SIGO from Run $2$vH$2$, including that of its shock front. Mach numbers are labelled based on the local temperature and the velocity relative to the center of mass of the SIGO. The SIGO itself here has a maximum radius of $118$ pc, which is slightly smaller than the sonic scale. The volume-averaged Mach number within the SIGO is $0.97$. One way to think about this SIGO is as a density fluctuation induced by this supersonic turbulence: the strongest density perturbations from the supersonic motions take place at the sonic scale, appearing as gas overdensities whose positions are correlated with those of nearby dark matter overdensities. This creates gas objects (SIGOs) offset from their parent halos, with sizes that are by necessity comparable to the sonic scale. Once this turbulence-induced structure formation occurs, molecular cooling helps to cool the gas within the SIGO to temperatures of order $200$ K \citep{Nakazoto+22}, which enables further collapse. Collapse of the SIGO reduces its axis length from the sonic scale, so a typical SIGO will have a maximum axis length that is comparable to but smaller than the sonic scale, unless it either accretes additional gas from its surroundings (permitting it to grow while accreting mass) or collapses.

\begin{figure}[t]
\centering

\gridline{\fig{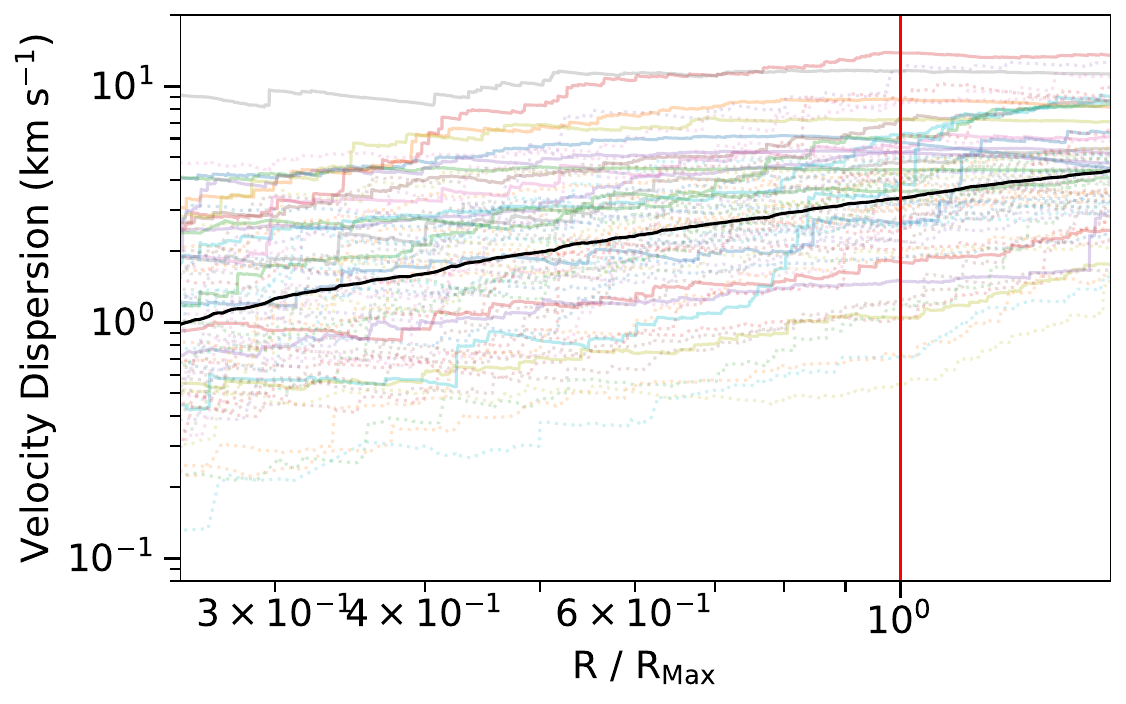}{0.46\textwidth}{Velocity Dispersion}
          }
\gridline{\fig{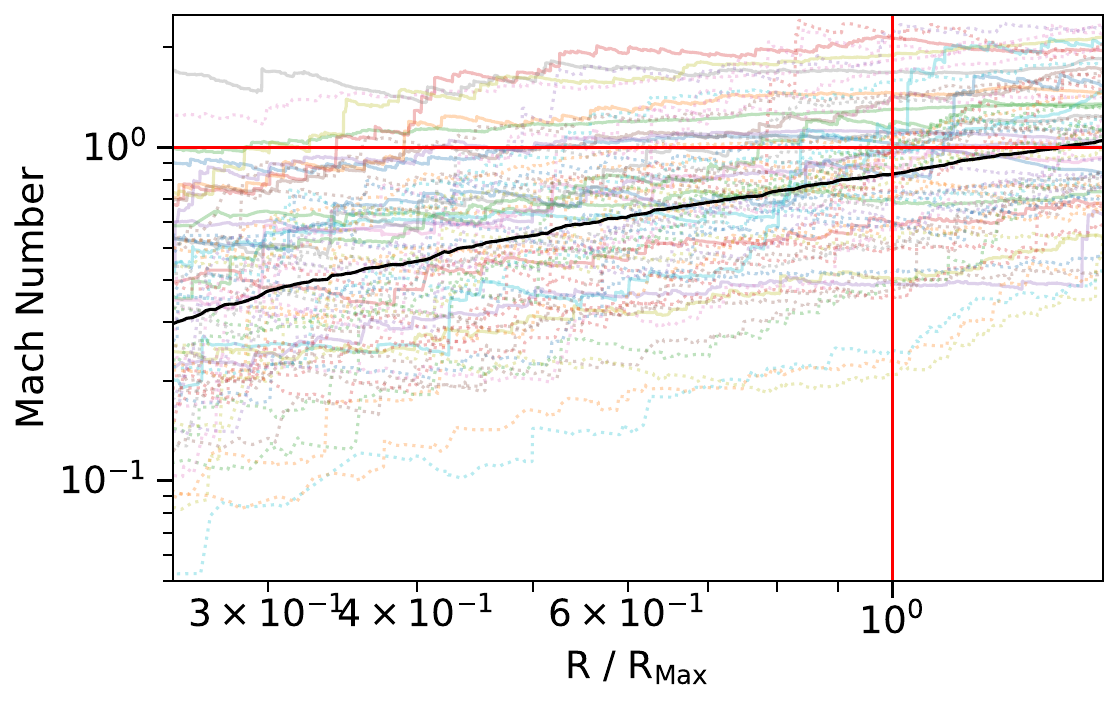}{0.46\textwidth}{Mach Dispersion}
          }
\caption{\textbf{SIGOs' linewidth-size relation: } Panel (a) shows the linewidth-size relation for SIGOs, computed over a sphere centered on the center of mass of each SIGO, with increasing radius (scaled for each SIGO by the length of the longest axis of that SIGO). The translucent lines represent individual SIGOs, and the black opaque line represents an average of all SIGOs in log space. Dotted lines represent low-mass SIGOs below $10^5$ M$_\odot$, which are more affected by $2$-body relaxation resulting from our limited resolution (see text). Panel (b) shows the mach number dispersion, showing that most SIGOs are somewhat smaller than the sonic scale. The vertical red line in each panel displays the length of the longest axis of the SIGO, and the horizontal red line in panel b indicates Mach 1. }\label{fig:VelocitySphericalAverage}
\end{figure}

Figure~\ref{fig:VelocitySphericalAverage} shows a more general comparison of the sonic scales and the maximum axis lengths of the SIGOs in Run $2$vH$2$. In the top panel, each faded line shows the velocity dispersion within a SIGO as a function of its radius, and in the bottom panel each faded line shows the mach number dispersion as a function of radius. The black line is the average of all SIGOs' dispersions in log space. The spatial distributions of each are similar, because SIGOs are close to isothermal: their sound speeds do not change much on the edges of the SIGOs compared to the centers. Typical temperatures at the edge of a SIGO are of order $10\%$ higher than those in the central regions. Therefore, their mach dispersions roughly follow their velocity dispersions. 

A typical SIGO has a volume-averaged velocity dispersion of about Mach 0.8, or about $3$ km/s (with fairly high variance) on the scale of its largest axis. Put another way, a typical SIGO has a sonic scale that is about $1.6$ times larger than its longest axis. This assumes a driving scale for the turbulence on the order of the maximum axis length of the SIGO, as discussed by \citet{chiou+19}. Because the sonic scale is comparable to the size of a SIGO (and in some cases is smaller than the SIGO), it is important to account for this turbulence when considering the ability of a SIGO to undergo gravitational collapse
(see also \cite{Hirano17} for similar effect in non-SIGOs). \citet{chiou+19} accounted for this effect by adapting a critical density \citep[previously used in GMCs by e.g.,][]{Krumholz+05}, defined by Equation (\ref{Eq:rhocrit}). This critical density defines the scale on which turbulence-induced fluctuations collapse through Jeans collapse, in effect equating the Jeans scale to the sonic scale. However, the analysis in the paper treated SIGOs as spherical objects with uniform density $\rho_{\rm SIGO}$ within a radius ${\rm R}_{\rm max}$ defined as the longest principal axis of the SIGO. SIGOs are not spherical (particularly with molecular hydrogen cooling accounted for), and have a typical shortest axis length R$_{\rm min}$ that is of order $10$ times smaller than their longest axis in Run $2$vH$2$; therefore, this overestimates SIGOs' masses. An updated treatment for the critical density of collapse for SIGOs, then, equates the mass of the SIGO to the Jeans mass, giving a new definition of the critical density of a SIGO for collapse $\rho_{\rm crit}$:
\begin{equation}\label{Eq:newrhocrit}
\rho_{\rm crit} \approx \frac{\pi c_s^2 \mathcal{M}^4} {4 G {\rm R}_{\rm min} {\rm R}_{\rm int}},
\end{equation}
where R$_{\rm int}$ is the second-largest principal axis of the SIGO and where we have assumed L$_{\rm drive}\approx 2\times {\rm R}_{\rm max}$.

At $z=20$, no SIGOs in Run $2$vH exceed this critical density. This result contrasts with that of \citet{chiou+19}, which found that SIGOs can collapse through only atomic hydrogen cooling, because of the additional factor of the prolateness of the SIGOs considered in this analysis. With atomic cooling alone, the cooling timescale for these SIGOs is long, so it is unlikely that SIGOs could collapse to potentially form stars outside of halos without considering molecular hydrogen cooling. However, when considering molecular hydrogen cooling, looking only at these SIGOs' densities at early redshifts does not paint the whole picture of SIGOs' ability to form stars outside of halos.

While the critical density and the Jeans density are both important to SIGOs, their location outside nearby halos allows their evolution to be characteristically slow. At $z=20$, about $50$ Myr or less after SIGOs begin to form in meaningful quantities, SIGOs tend to be fairly isolated objects. They are outside of the immediate vicinity of the DM halos that birthed them, and have characteristically long fall-back times to their nearest halo. Because of this, even though their cooling times are long \citep{Schauer+21}, SIGOs can have shorter cooling times than fall-back times, allowing them to collapse in spite of their early Jeans stability (though typically later than $z=20$, because these timescales tend to be $\gtrapprox 50$ Myr). This may permit the formation of stars. Subsequently, they may be able to accrete onto halos on the fall-back timescale, and potentially survive as identifiable clusters due to their compact nature and existing population of stars. In order for this to happen, they must have cooling and collapse timescales that are shorter than their fall-back timescale to nearby halos, as discussed in Section~\ref{sec:evolution}. At any given time, then, SIGOs may be destined for collapse through cooling despite not exceeding the critical density for collapse. Therefore, to study the evolution of SIGOs, we argue that a timescale analysis is more appropriate than a density analysis, as follows in Section~\ref{sec:evolution}.

Another of the primary characteristics of GMC populations is their power law mass spectrum. GMCs in the inner disk of the Milky Way follow a power law mass spectrum given by dN$\propto$M$^\xi$dM with $\xi\sim-1.5$. In contrast, the mass spectra of GMCs in the outer Milky Way ($\xi\sim-2.1 \pm 0.2$) and M$33$ ($\xi\sim-2.9 \pm 0.4$) are notably steeper \citep[e.g.][]{Rosolowsky+05}. Characterizing this mass spectrum is vital to understanding both how the clouds formed, as well as their overall contribution to large-scale star formation.  Establishing how the mass spectrum of SIGOs compares to these power laws is useful not only for furthering theoretical work with SIGOs, but also for contextualizing similarities between SIGOs and these different populations of GMCs.

\begin{figure}[t]
    \centering
    \includegraphics[width=\linewidth]{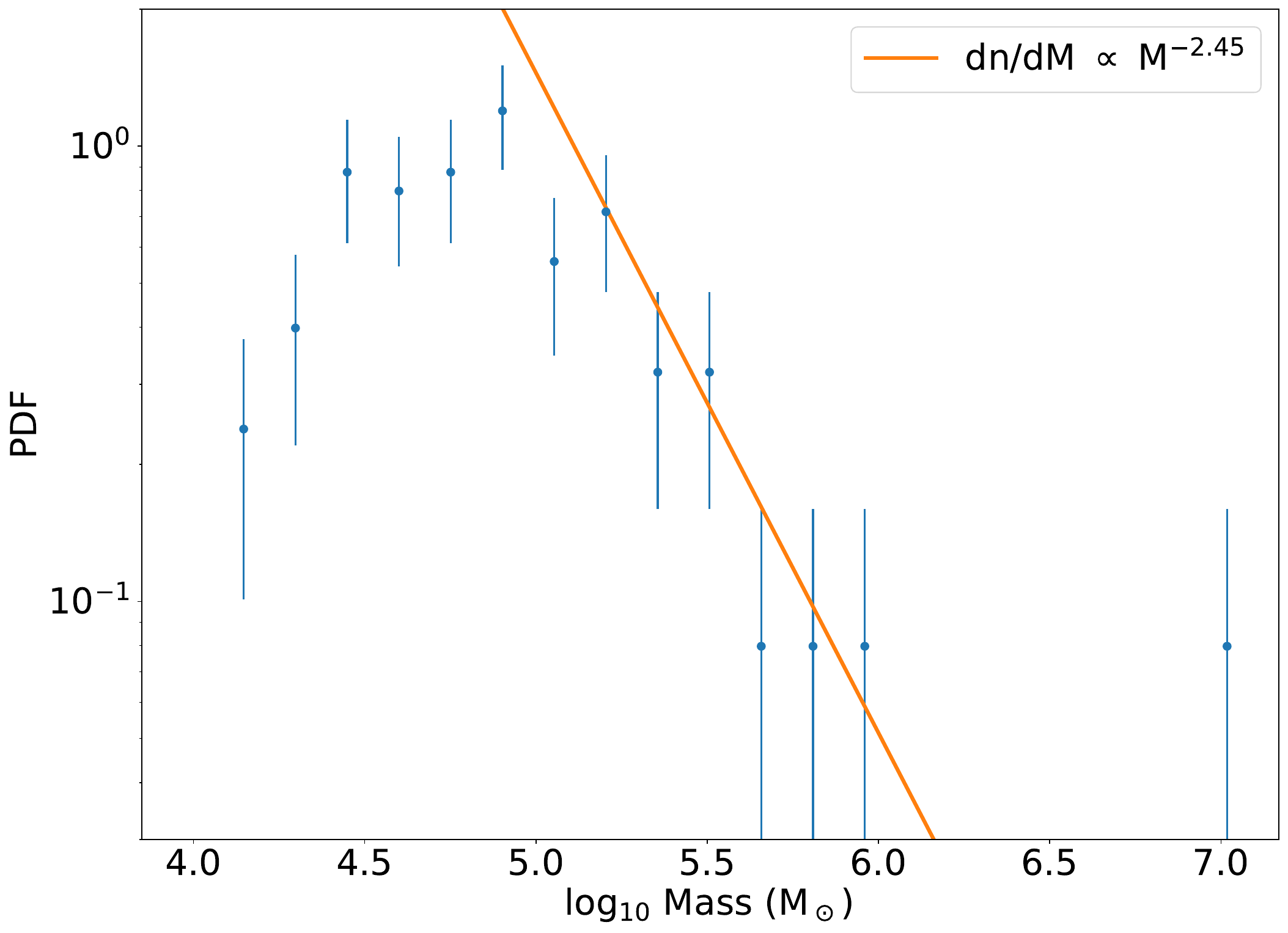}
    \caption{\textbf{The Mass Spectrum of SIGOs: } Here we show the mass spectrum of SIGOs from Run $2$vH$2$ at $z=20$ (the blue points and associated Poisson errors, plotted as a probability density in log space), as well as showing (in orange) a best-fit power law mass spectrum for high-mass SIGOs. The mass spectrum of SIGOs in this Run is consistent with a power law index of $-2.4\pm0.3$.}
    \label{fig:MassSpectrum}
\end{figure}

Figure~\ref{fig:MassSpectrum} shows the mass spectrum of SIGOs in Run $2$vH$2$ at $z=20$. The blue histogram points show the probability density of finding a given SIGO in the labelled mass bin, with Poisson errors. However, immediately translating this estimate of the mass spectrum to a power law is complicated by the artificial $2$-body relaxation present in the simulation (see Section~\ref{ssec:relaxation} for more details). At low masses, this mass spectrum is distorted, or truncated, by the limited resolution of our simulation. Therefore, we only use mass points above M$_{\rm SIGO,min}$ = $10^5$ M$_\odot$ ($278$ gas particles) to construct a best-fit power spectrum. This cutoff reflects SIGOs whose artificial $2$-body relaxation timescales are longer than the age of the oldest SIGO in the simulation, which are therefore less affected by relaxation. We introduce an upper bound mass cutoff of $10^6$ M$_\odot$, to avoid over-fitting to the SIGO with a mass of $\sim 10^7$ M$_\odot$, because we do not have enough data points to determine whether or not the spectral index varies over the range $10^6$ M$_\odot \leq$ M$_{\rm SIGO}$ $\leq 10^7$ M$_\odot$. The best-fit power law is expected to diverge from the data in this range due to Poisson fluctuations. We calculate a maximum likelihood power law index $\hat{\xi}$ using the method of \citet{Clauset+09} (derived from \citealt{Muniruzzaman+57}), using our lower-bound mass cutoff of $10^5$ M$_\odot$ as M$_{\rm SIGO,min}$ as follows:

\begin{equation}\label{Eq:BestFitXi}
\hat{\xi} = 1 + n\left[\sum_{i=1}^n {\rm ln}\frac{{\rm M}_{{\rm SIGO},i}}{{\rm M}_{\rm SIGO,min}} \right]^{-1},
\end{equation}
where $n$ is the number of SIGOs with $10^5$~M$_\odot \le$~M$_{\rm SIGO}\le 10^6$~M$_\odot$, and ${\rm M}_{{\rm SIGO},i}$ is the mass of each of these SIGOs.

The error $\sigma_\xi$ is then estimated from the width of the maximum likelihood estimate as 
\begin{equation}\label{Eq:MLEError}
\sigma_\xi = \frac{\hat{\xi} - 1}{\sqrt{n}} + {\rm O}\left(\frac{1}{n}\right).
\end{equation}

\begin{figure*}[t]

\centering
\includegraphics[width=\linewidth]{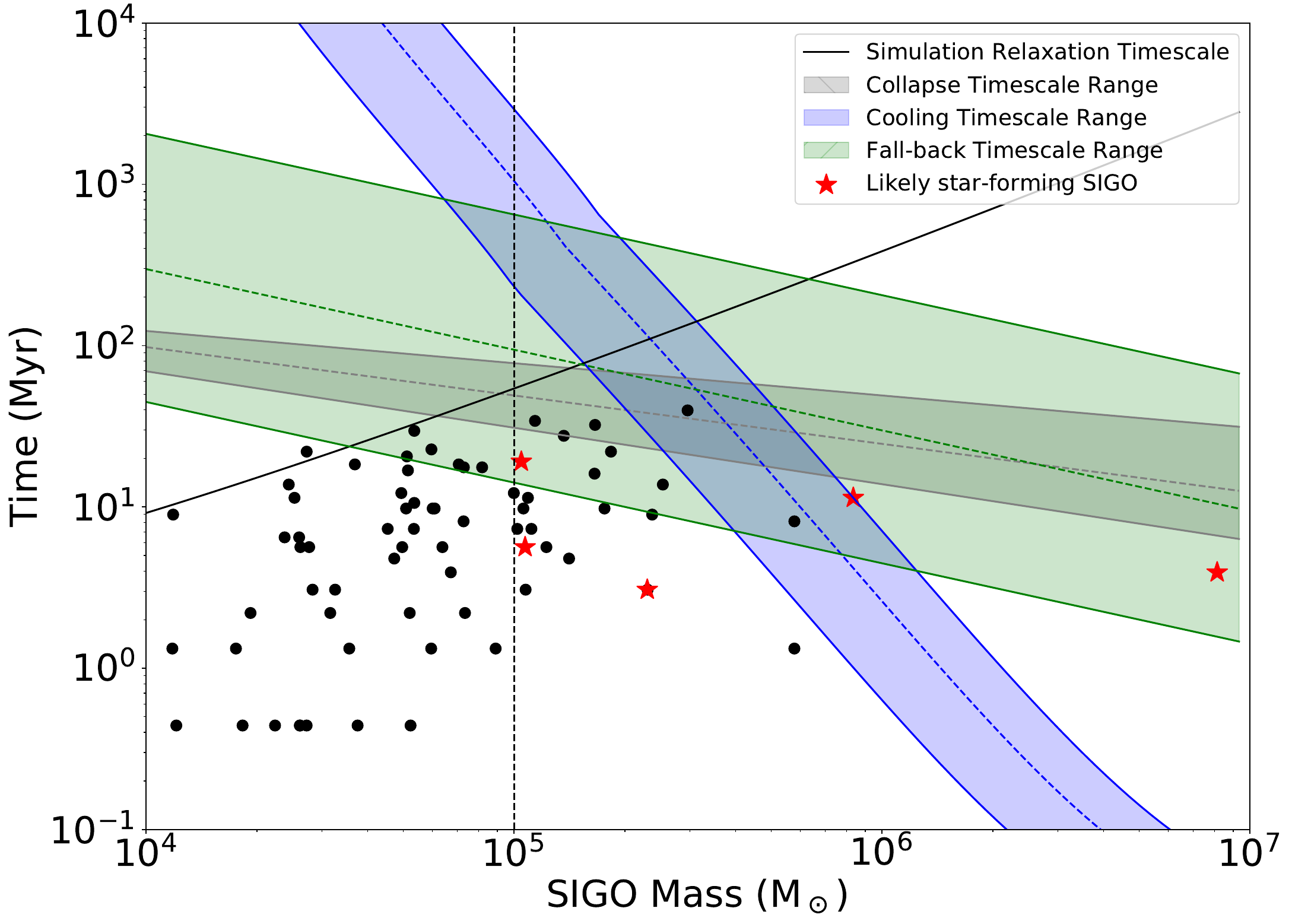} 
    
    \caption{{\bf Timescales of SIGOs:} here we show a number of important timescales to the evolution of SIGOs at $z=20$ in Run $2$vH$2$. Black dots mark the age and baryon mass of SIGOs found at this redshift, which represent $16\%$ of the SIGOs that form in our simulation at all redshifts. Ranges are representative of the properties of SIGOs at $z=20$ in our simulation (see text for details of assumptions). The green region and lines in these figures show the range of fall-back timescales to the nearest halo, as a function of mass. The green dashed line indicates the same for a SIGO with median properties. The blue region shows the same, but for the range of cooling timescales, and the gray region  shows the timescale for gravitational collapse of a SIGO. The black line shows the relaxation timescale of SIGOs in the simulation at the resolution of the simulation (much shorter than in the real Universe). The black dashed line indicates the mass scale above which simulated SIGOs are less affected by relaxation at $z=20$, with maximum ages shorter than their relaxation timescales. SIGOs with shorter cooling and collapse timescales than their fall-back timescales are marked with red stars as having the potential to form stars outside of a DM halo. As depicted, the main limitation on low mass SIGOs' lifetimes is the numerical evaporation process (i.e., 2-body relaxation, see text). Therefore, in our adopted resolution, high-mass SIGOs can collapse to potentially form stars, while in the Universe we expect that more SIGOs will form stars.   }\label{Fig:AgeVsMass}
\end{figure*}

With these conditions in place, the orange line in Figure~\ref{fig:MassSpectrum} shows a maximum likelihood power law for our simulated SIGOs' masses. We find a best-fit mass spectrum index for SIGOs in this Run of $\xi\sim-2.4\pm0.3$, which is, interestingly, consistent with the mass spectrum of GMCs in the outer Milky Way, and is broadly consistent with the range of mass spectra seen in GMC populations, despite their differing formation routes.

\section{Timescale Analysis of the Evolution of SIGOs}\label{sec:evolution}

The similarity between SIGOs and GMCs is limited when considering the main challenge for SIGOs: their location outside of a DM halo, and eventual fall back into a DM halo. In considering this challenge, and as a starting point for our timescale analysis, we identify the following evolutionary channels for SIGOs in a simulation (see Figure~\ref{Fig:AgeVsMass} for the relevant processes). 
\begin{itemize}
    \item {\it Gravitational Collapse}: SIGOs can undergo gravitational collapse in their overdense cores, or on the scale of the SIGO. We expand on this physical process in \S \ref{ssec:Collapse}. 
    \item {\it Cooling}: As mentioned above, molecular hydrogen cooling lowers the temperature of the gas, reducing the gas pressure and assisting with collapse. See \S \ref{ssec:cooling} for more details.  
    \item {\it Fall-back Into Halos}: SIGOs are inherently found near DM halos \citep{naoznarayan14}. Therefore, SIGOs are likely to eventually fall into a DM halo. Accretion is the most likely final state of even potential star-forming SIGOs, and could lead to the formation of accreted clusters. See \S \ref{ssec:freefall} for an estimate of this timescale.
    \item {\it $2$-body Relaxation:} Gas in {\tt AREPO} numerical simulations has an associated mass that depends on the simulation resolution. This gas interacts gravitationally, causing it to undergo artificial two-body relaxation processes. These artificial interactions result in the evaporation of gas structures on a characteristic timescale. We estimate the timescale of this process in \S \ref{ssec:relaxation}. 
    \item {\it Growth}: SIGOs can grow by accreting gas and dark matter from their environment. Over time, this changes their mass and gas fraction, and can impact their ability to collapse. We overview this process in \S \ref{ssec:accretion}
\end{itemize}


\subsection{Gravitational Collapse}\label{ssec:Collapse}

The timescale for gravitational collapse is important to determining whether a SIGO can achieve the overdensities needed to form stars. In order for this to occur, a SIGO needs to be able to collapse within a shorter time span than the fall-back timescale to nearby DM halos. If this condition is not satisfied, the SIGO will fall into a nearby DM halo before it forms stars, becoming disrupted by tides or external pressure. As a starting point for this timescale comparison, we calculate a free collapse timescale:
\begin{equation}\label{Eq:collapsetime}
t_{\rm collapse} \approx \frac{1} {\sqrt{G\rho_{\rm peak}}} \ ,
\end{equation}
depicted in Figure~\ref{Fig:AgeVsMass} for a representative SIGO, and in  Appendix \ref{Sec:appendix}, Figure~\ref{Fig:CollapseTimescale} for a depiction of individual SIGOs. 
For the collapse timescale for a typical SIGO  in Figure~\ref{Fig:AgeVsMass}, we assume a representative range of peak densities, based on the range of densities in SIGOs at $z=20$ in Run $2$vH$2$ (see Appendix \ref{Sec:appendix} for more details). The upper bound density is taken as 
\begin{equation}\label{Eq:MaxDensity}
{\rm log}_{10}(\rho_{\rm max,com}) = 0.7\times {\rm log}_{10}\left(\frac{{\rm M}_{\rm SIGO}}{{\rm M}_\odot}\right) - 30.2. 
\end{equation}
The lower bound density is taken to be
\begin{equation}\label{Eq:MinDensity}
{\rm log}_{10}(\rho_{\rm min,com}) = 0.4\times {\rm log}_{10}\left(\frac{{\rm M}_{\rm SIGO}}{{\rm M}_\odot}\right) - 29.6,
\end{equation}
where $\rho$ has units of g/cm$^3$.
Here we use peak rather than mean densities to reflect that the SIGO's star formation is likely primarily occurring in its most dense regions: the process of star formation in a SIGO is not $100\%$ efficient. 

As seen in Figure~\ref{Fig:AgeVsMass}, the collapse timescale has a negative dependence on mass, driven by larger central overdensities in more massive SIGOs. This mass dependence contributes to the enhanced ability of massive SIGOs to form stars. 


\subsection{Cooling}\label{ssec:cooling}

However, this timescale for free gravitational collapse does not paint the whole picture of the collapse of SIGOs. In order for a SIGO to collapse, it also must be able to efficiently cool, allowing its Jeans mass to decline and permitting gravity to overcome gas pressure. There are 2 phases to this cooling in a SIGO that has an initial Jeans mass above its actual mass: in the first phase, the SIGO isochorically cools, lowering its Jeans mass until it drops below the SIGO's mass. Secondly, the SIGO begins to collapse, with the dynamics of its collapse determined by the cooling and collapse timescales. To account for this, we can define a cooling timescale ${\rm t}_{\rm cool}$, given as the time it would take for a gas clump to cool isochorically at a rate $\Lambda(T,n_{\rm H})$ to its Jeans temperature, or, for SIGOs with supersonic velocity dispersions, to a temperature at which its density exceeds the critical density. We then add the timescale for the second phase of this collapse: the cooling time as defined in \citet{Schauer+21} and derived from \citet{Hollenbach+79}, taken from the temperature at which the SIGO initially collapses.


\begin{equation}\label{Eq:coolingtime}
t_{\rm cool} = \frac{k_{\rm B}T_{\rm collapse}}{n_{\rm H}(\gamma-1)\Lambda(T_{\rm collapse})} + t_{\rm iso},
\end{equation}
where $\gamma$ is the adiabatic index and t$_{\rm iso}$ is the additional cooling time a SIGO that initially does not exceed its Jeans mass takes to isochorically cool to lower its Jeans mass to M$_{\rm SIGO}$ and begin to collapse:
{\large{
\begin{equation}\label{Eq:t_iso}
{\rm t}_{\rm iso} = \left\{ 
  \begin{array}{ c l }
    0 & \quad {\scriptstyle {\rm M}_{\rm SIGO} \geq  {\rm M}_{\rm J} ({\rm T}_{\rm init})} \\
    & \\
   \int_{{\rm T}_{\rm Collapse}}^{{\rm T}_{\rm init}}  \frac{k_{\rm B}}{n_{\rm H}(\gamma-1)\Lambda(T)} dT      \ .     & \quad {\scriptstyle {\rm M}_{\rm SIGO} < {\rm M}_{\rm J} ({\rm T}_{\rm init})}
  \end{array}
\right.
\end{equation}
}}
If the SIGO initially exceeds the Jeans mass, its collapse temperature is taken as its volume-averaged temperature. Otherwise, its collapse temperature (after the SIGO has isochorically cooled), is taken as the Jeans temperature that the SIGO has cooled to:

\begin{equation}\label{Eq:T_Jeans}
{\rm T}_{\rm J} = \frac{3{\rm G}\mu {\rm m}_{\rm H}}{5{\rm k}_{\rm B}} \sqrt[3]{\frac{36 {\rm M}_{\rm SIGO}^2 \rho_{\rm SIGO}}{\pi^2}}
\end{equation}

In SIGOs, $\Lambda$(T) is dominated by and approximated as the molecular cooling rate defined in \citet{Galli+98}:
\begin{equation}\label{Eq:lambda}
\Lambda({\rm T})_{\rm H_2} = \frac{\Lambda({\rm LTE})}{1+[n^{\rm cr} / n({\rm H})]},
\end{equation}
consistent with the molecular cooling prescription used in GRACKLE.
Note here that because the SIGO may accrete gas from its surroundings and gain mass as it cools (raising its Jeans temperature), because the SIGO also contracts as it cools on a similar timescale \citep{Nakazoto+22}, and because this does not account for overdense regions within the SIGO which enhance cooling, this cooling timescale assuming an isochoric process is an upper bound on the true cooling time relevant to collapse.

We show the representative timescale based on Equation~\ref{Eq:t_iso}
in Figure~\ref{Fig:AgeVsMass} (see also Figure~\ref{Fig:CollapseTimescale} from Appendix~\ref{Sec:appendix}, for this timescale for individual SIGOs ). To depict the typical value of this timescale we take the median initial gas number density of a SIGO, $1.08$~cm$^{-3}$. To represent the maximum cooling time, we take a lower bound gas density of $0.7$~cm$^{-3}$. For the lower bound cooling time, we take n$_{\rm H} \sim 2.0$~cm$^{-3}$ (see Figure~\ref{Fig:Densities} from Appendix \ref{Sec:appendix}). We assume an initial temperature of $500~{\rm K}$ or the Jeans temperature (Eq \ref{Eq:T_Jeans}), whichever is higher, for this representative line. 
As depicted in Figure~\ref{Fig:AgeVsMass}, most SIGOs at $z=20$ have not yet cooled substantially. 

\subsection{Fall-back Into Halos}\label{ssec:freefall}

\begin{figure}[t]

\centering
\includegraphics[width=\linewidth]{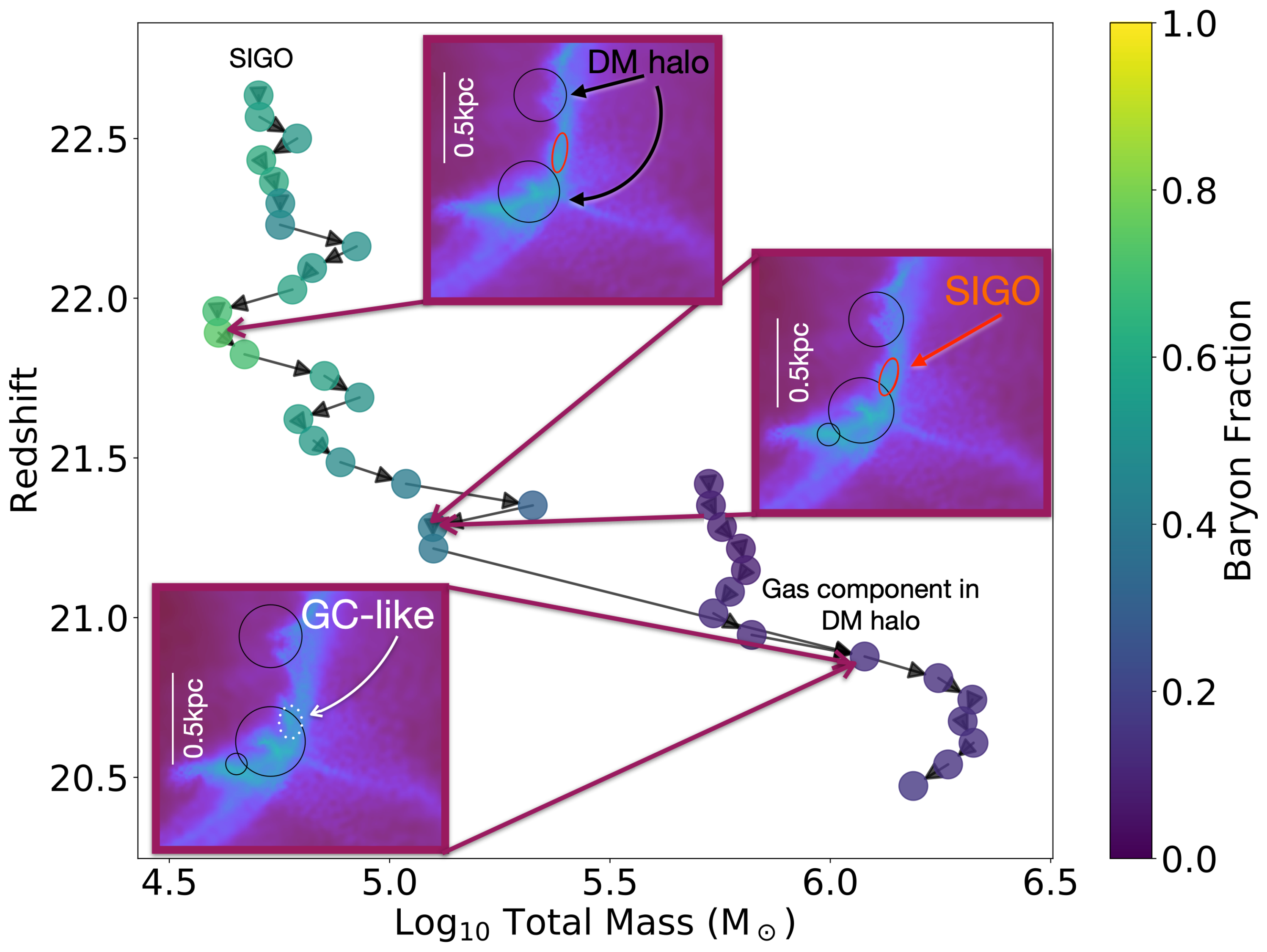} 
    
    \caption{{\bf Fall-back of a SIGO (the creation of a GC-like system at high redshift):} here we show the time evolution of a single SIGO in Run $2$vH$2$. This SIGO, the left-most branch of this plot, is accreted by a nearby halo (the gas component of which is shown in purple, the right branch of the plot, with a very low baryon fraction). This accretion produces a more massive merged object, containing the SIGO in its substructure. This SIGO (labelled ``GC-like" in the figure, for globular cluster-like candidate), will be subject to future evolution within the halo, and is identifiable as a distinct component, with the potential to continue to evolve into a star cluster. For a movie of this evolution see \href{https://www.astro.ucla.edu/~snaoz/TheSupersonicProject/images/SIGOsGC-like.mp4}{here}.
    }\label{Fig:SIGOFallback}
\end{figure}

Eventually, the majority of SIGOs will be accreted onto a halo (most often their parent halo), forming globular-cluster-like systems. Like GCs, these are mostly expected to reside in the halos of their host galaxies \citep[e.g.,][]{Chaboyer+99,Benedict+02,Chen+18}. \citet{naoznarayan14} calculated the timescale on which a SIGO free-falls to its parent halo:
\begin{equation}\label{Eq:freefalltime}
t_{\rm ff} = 0.27 {\rm Gyr} \left(\frac{\Delta r_{\rm phys}}{0.59 {\rm kpc}}\right)^\frac{3}{2} \left(\frac{\Omega_m}{\Omega_b} \frac{{\rm M}_{\rm b}}{10^6 {\rm M}_\odot}\right)^{-\frac{1}{2}},
\end{equation}
where $\Delta r_{\rm phys}$ is the physical separation between the SIGO and the halo, and ${\rm M}_{\rm b}$ is the baryon mass of the SIGO.
In order for a SIGO to collapse and potentially form stars outside of a DM halo, this timescale must be longer than both the cooling and collapse timescales for that SIGO, permitting the object to cool and reach high central densities prior to being affected by tides or ram pressure stripping near the larger halo. 

We show the representative timescale from Eq.~(\ref{Eq:freefalltime}) in Figure~\ref{Fig:AgeVsMass}, where we assumed a median physical separation of $0.4$ kpc between a SIGO and its nearest halo, with a range of $ 0.1-0.9$ kpc for a characteristic SIGO, representative of the true range of separations in the simulation at $z=20$. As seen in this figure, some SIGOs ``fall'' into a DM halo. Figure~\ref{Fig:SIGOFallback} shows the evolutionary path of a single such SIGO (left of the figure with high baryon fractions) that falls back into a dark matter halo (the gas component of which forms the right branch of the figure, with low gas fractions). Notably, this particular SIGO was accreted by a larger nearby DM halo rather than its parent halo, allowing its lifetime outside of a halo to be unusually short for this process, only $15$ Myr. We label the final outcome as ``GC-like,'' indicating the potential for this object to form a star cluster at the outskirts of the halo, similar to a present-day GC.

\subsection{2-body Relaxation}\label{ssec:relaxation}

In a simulation with limited resolution, each Voronoi cell has an associated gas mass. Therefore, gravitational interactions between these cells can lead to changing the object's energy by an order of itself. This processes, called $2$-body relaxation, has been shown to be a possible limiting factor on simulations of SIGOs \citep{naoznarayan14,popa}. This process essentially ``evaporates'' the object as a function of time. The $2$-body relaxation timescale for $N$ particles can be written as 

\begin{equation}\label{Eq:relaxtime}
t_{\rm rlx} \sim \frac{{0.1 \rm N}}{ {\rm ln}({\rm N})} \times \frac{{\rm r}}{\sigma},
\end{equation}
\citep[]{BinneyTremaine} where $\sigma$ is the $1$D velocity dispersion within the SIGO and $r$ is taken to be the minimum principle axis length (ergo ${\rm r}/\sigma$ is t$_{\rm cross}$, such that this relaxation time describes the timescale on which the SIGO significantly changes in size). SIGOs can have as few as $\sim 100$ gas particles in these simulations, which combined with their small sizes and fairly high velocity dispersions leads to artificial relaxation times on the order of tens of Myr. This underscores the need for caution when following the evolution of SIGOs to low redshifts in these cosmological simulations: at our resolution, low-mass SIGOs with around $10^4$ M$_\odot$ of material are already beginning to be destroyed by artificial relaxation at redshift $20$.

We depict the timescale from Eq.~(\ref{Eq:relaxtime}) in Figure~\ref{Fig:AgeVsMass} (see also Figure~\ref{Fig:CollapseTimescale}), as a solid black line. As seen in the Figure, the main effect shortening the lifetime of our SIGOs is the relaxation timescale generated by our limited resolution. Lower-mass SIGOs left of the black dashed line at $10^5$ M$_\odot$ may be destroyed by this relaxation before $z=20$. Thus, we expect SIGOs in the Universe to have increased longevity not captured by our simulation. 

\subsection{Growth}\label{ssec:accretion}

\begin{figure}[t]

\centering
\includegraphics[width=\linewidth]{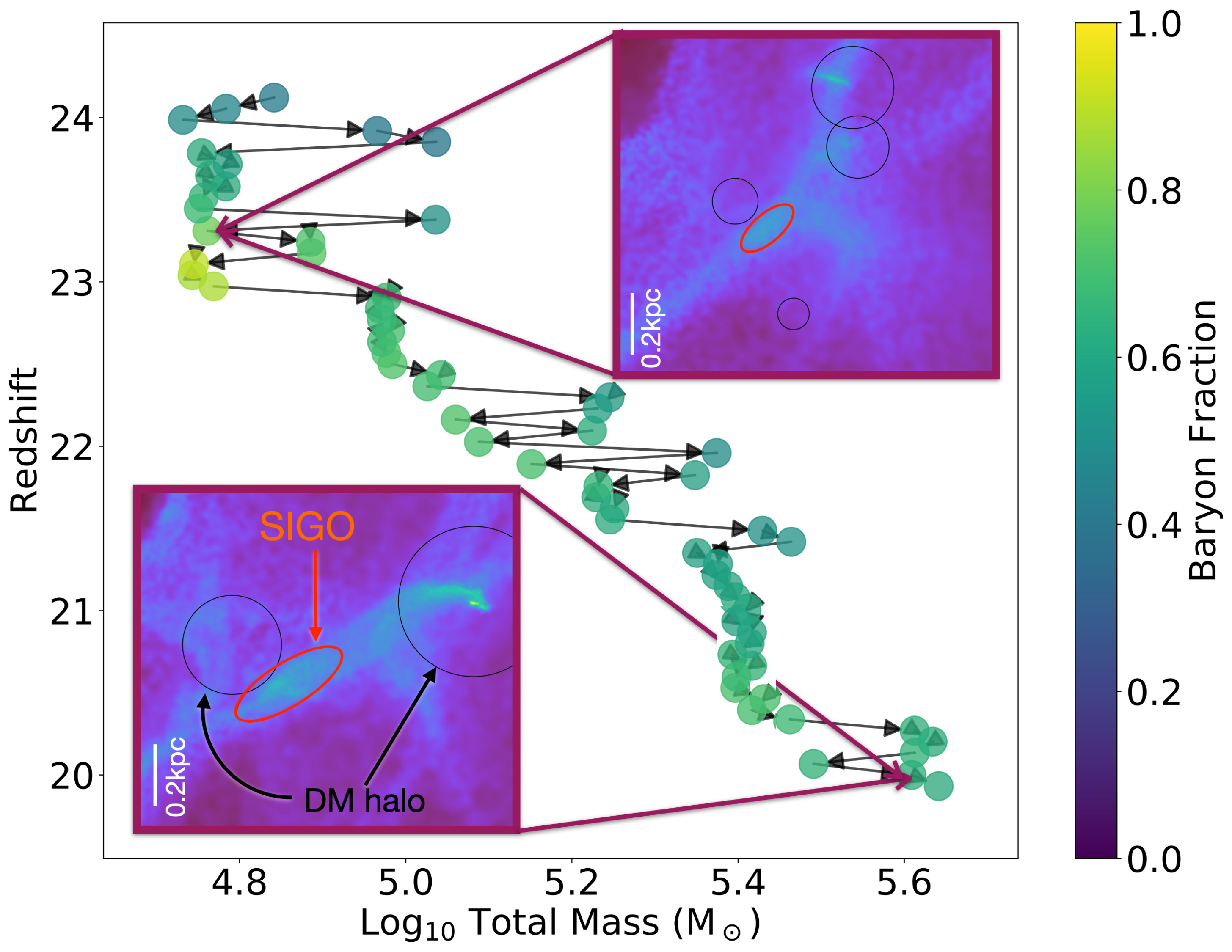} 
    
    \caption{{\bf Growth of a SIGO:} here we show the time evolution of a single SIGO in Run $2$vH$2$. This SIGO increases its mass by nearly a factor of $10$ between its formation and the final redshift of our simulation, accreting gas from its surroundings with a similar baryon fraction to the SIGO itself and thereby maintaining its baryon fraction over time. For a movie of this evolution see \href{https://www.astro.ucla.edu/~snaoz/TheSupersonicProject/images/SIGOGrowth.mp4}{here}.}\label{Fig:SIGOGrowth}
\end{figure}

While constraining the full spectrum of growth timescales is challenging at present due to limitations of our spatial and time resolution, it is important to also note that SIGOs can grow through gas and dark matter accretion. As can be seen in Figure~\ref{fig:ExampleSIGOMach}, SIGOs form embedded in gas streams--typical gas densities surrounding the forming SIGOs can be as high as $\sim 20\%$ that of the SIGO itself. In addition, the material around the SIGO may be itself enriched in baryons (though its gas fraction will not necessarily be as high as that of the SIGO). This gas, as well as some dark matter, may be accreted from a SIGO's environment, increasing its mass over time and allowing its gas fraction to change. 

Figure~\ref{Fig:SIGOGrowth} shows an example of this growth for a single SIGO in Run $2$vH$2$. This SIGO formed with a mass of about $6\times10^4$ M$_\odot$ and a gas fraction near $60\%$. The SIGO was able to accrete gas from its surroundings with a similar baryon fraction to the SIGO itself, growing by a factor of nearly 10 while maintaining a roughly constant gas fraction. This has the potential to impact the SIGO's future evolution, as the more massive SIGO at $z=20$ is much more likely to quickly cool and collapse to form stars through the process outlined above in \S \ref{ssec:cooling}. 

It is important to note, however, that this process of growth does not have to maintain a SIGO's gas fraction. Over time, accretion of surrounding material can cause the gas fraction in SIGOs to drop below the $60\%$ threshold set for SIGO identification, even though the SIGOs still exist (and are commonly still enriched in gas). In Run $2$vH$2$, about $50\%$ of formed SIGOs at all redshifts were found to eventually drop below a gas fraction of $60\%$ through this process before $z=20$, while generally maintaining a gas fraction above $40\%$ and maintaining or increasing their density. In future studies modelling star formation in SIGOs to lower redshifts, it will be important to follow these evolved SIGOs, which have higher masses than newborn SIGOs and may be promising grounds for star formation in SIGOs at lower redshifts.

\subsection{Timescale Comparison}

Figure~\ref{Fig:AgeVsMass} shows these characteristic timescales for the evolutionary channels for SIGOs mentioned above. In grey, one can see the characteristic timescale for gravitational collapse of SIGOs identified at $z=20$ and at least one other snapshot\footnote{Here we use multiple snapshots in order to limit false detections of SIGOs and better constrain their dynamical properties.} of Run $2$vH$2$. In green, we show the range of fall-back timescales to the nearest halo in the simulation. In blue, we show the range of cooling times for the SIGOs to cool through molecular hydrogen cooling to Jeans collapse (for SIGOs with a Mach dispersion less than 1) or collapse through the critical density (for SIGOs with a Mach dispersion greater than 1). For all of these regions, we also add a dotted line, showing the timescale for a SIGO of a given mass with median properties (principle axis lengths, distance from the nearest halo, temperature, gas fraction, and velocity dispersion). We also add a black line showing a typical timescale for 2-body relaxation in the simulation at simulation resolution. The figure also shows the age and mass of each SIGO in Run $2$vH$2$ at $z=20$. The age is computed by comparing the first redshift at which each object meets the conditions to be identified as a SIGO (outlined in Section~\ref{sec:Meth}) to the time of the plot at $z=20$. Note that SIGOs with an age of $0.45$ Myr are those that have just formed in the simulation. The age is calculated by assuming that their formation time is halfway between the penultimate and final snapshots.

As one can see from the plot, the cooling timescale has a strong (roughly M$^{-2.6}$) dependence on mass. Because of this, molecular cooling alone is not sufficient to efficiently cool SIGOs below about $\sim 10^5$ M$_\odot$ to collapse. In essentially all cases with such SIGOs, the cooling timescale will be dramatically longer than the collapse and fall-back timescales, so the SIGO will fall into a halo before forming stars.

Above this mass limit, one must also compare the timescales for a SIGO to fall back into a nearby halo (which may be its parent halo or another gravitationally dominant nearby halo) or collapse gravitationally to determine SIGOs' fates. In isolation, high-mass SIGOs are frequently capable of collapsing to form stars; however, because of the wide variance of these timescales, it is possible for the fall-back timescale for a SIGO to be shorter than the collapse timescale. Because of this, the fall-back and collapse timescales must be individually compared, to determine which SIGOs will form stars before accretion onto nearby DM objects. Such SIGOs are marked in the plot with red stars.

One final consideration in studying the evolution of SIGOs apparent from this plot is the simulation's relaxation timescale. At the resolution of this study, even $10^5 $ M$_\odot$ SIGOs dissipate due to $2$-body relaxation within $80$ Myr--a similar length of time to that between the first SIGO's formation and our $z=20$ snapshot. Because of this relaxation, we are undercounting the lowest-mass SIGOs at $z=20$, as well as potentially underestimating the densities of those that exist. In order to eventually follow SIGOs through the process of forming stars and merging with halos at later times, future studies will need to work at higher resolutions ($\sim 200$ ${\rm M}_\odot$ or better), increasing the relaxation timescale.

\section{Discussion}\label{Sec:Discussion}

Understanding the behavior of supersonic turbulence in and around SIGOs is key to understanding their evolution and potential for star formation \citep{chiou+19}. Here we show how  supersonic turbulence acts together with molecular cooling in SIGOs to form density peaks, with high enough densities to become sites of star formation. This process is similar to that in GMCs, where turbulence has a scale-dependent effect on star formation, boosting densities on  the sonic scale \citep[e.g.,][]{Krumholz+05, burkhart18a}. 

The structure and population-level properties of SIGOs are also similar to those of GMCs. For example, we find that SIGOs have similar substructures and masses to GMCs. In particular, we find that SIGOs have core-like structures and filaments (Figure~\ref{Fig:Densities}, or see the evolved SIGO in Figure~\ref{Fig:SIGOGrowth}),  similar to those in GMCs \citep{MacLow+04,Krumholz+05,Mocz+18}. 
Further, we find that the mass spectrum of SIGOs is consistent with the mass spectrum of GMCs in the outer Milky Way (see Figure~\ref{fig:MassSpectrum} in \S \ref{Sec:Results}).

In particular, we show for the first time in a simulation that SIGOs form on scales comparable to the sonic scale (Figure~\ref{fig:VelocitySphericalAverage}). Supersonic turbulence aid in the formation and collapse of SIGOs, with SIGOs forming as high density peaks. Molecular cooling lowers these objects' Jeans scale. When this process is sufficiently efficient, this lowers the Jeans scale below the sonic scale and permits collapse of these newly formed SIGOs. This may permit star formation in SIGOs.

\citet{chiou+19} found that most SIGOs should form stars through atomic cooling alone. However, this result was called into question by \citet{Schauer+21}, which included atomic and molecular cooling and did not find star formation sites outside of DM halos in a simulation. \citet{Nakazoto+22}, on the other hand, followed a SIGO in a zoom-in simulation and confirmed that the SIGO experienced Jeans collapse. The modifications in this paper to the prescriptions in \citet{chiou+19} are a step towards reconciling these disparate results. Correcting the prescription for collapse in SIGOs based on their asphericity, we come to the conclusion that SIGOs should not form stars when considering only atomic hydrogen cooling. With molecular hydrogen cooling accounted for, SIGOs should be able to form stars. However, not all SIGOs undergo global collapse, exceeding the Jeans scale on the physical scale of the entire SIGO rather than in substructure \citep[neglecting metal line cooling, which has the potential to enhance star formation in SIGOs but has not yet been studied in a simulation; see e.g. ][]{Schauer+21}. 

At $z=20$, when most SIGOs have not yet cooled, about $5$ SIGOs will be capable of undergoing Jeans collapse. Thus, at lower redshifts, we expect many SIGOs to collapse. 
We estimate a star-forming SIGO abundance of $0.63$ Mpc$^{-3}$ at $z=20$.
This abundance is comparable to the present-day local density of low-metallicity globular clusters (i.e., $0.44$ Mpc$^{-3}$, \citealt{Rodriguez+15}). Because of the high Poisson uncertainty inherent to this figure, this suggests an explanation for the lack of star-forming SIGOs in \citet{Schauer+21}. A smaller simulation box size and $\sigma_8$, combined with lower star-forming SIGO abundances than could be inferred from the literature at the time, may allow a $<1.5 \sigma$ Poisson fluctuation to yield a simulation box with no star-forming SIGOs outside of halos, neglecting metal line cooling. However, these results indicate that the abundance of star-forming SIGOs in the Universe may be quite high in regions with significant stream velocities. Because our local neighbourhood has an estimated $1.75 \sigma$ stream velocity fluctuation \citep{Uysal+22}, which is $87.5\%$ of the value of the stream velocity studied in this work, this is potentially an important contributor to the early structure of our own Local Group.

The effect of $\sigma_8$ on the redshift of SIGO formation must also be considered in contextualizing these results. Here, we have assumed an elevated $\sigma_8$, representative of a $\sim 2\sigma$ density fluctuation. This elevated value of $\sigma_8$ yields more power; i.e., it helps structure develop earlier \citep[e.g.,][]{greif11,park+20}. As shown in Figure~\ref{Fig:Densities} (see Appendix \ref{Sec:appendix}), the comoving density of SIGOs before they collapse is not significantly dependent on redshift, so the pre-collapse physical density and therefore the cooling rate of SIGOs is dependent on their redshift of formation. In overdense regions such as those that form galaxy clusters, then, SIGOs can more efficiently cool and collapse than in underdense regions. This factor explains why SIGOs are not seen in regions of the Universe outside of galaxies: SIGOs formed too late to efficiently cool in the underdense regions of the early Universe that gave rise to these volumes. On the other hand, SIGOs form earlier and are more prone to form star clusters in overdense regions.

As highlighted in Figure~\ref{Fig:AgeVsMass}, the resolution limitation yields an artificial age limit on SIGOs, which should not exist in the case of gas objects\citep[e.g.,][]{naoznarayan14}. Thus, we use a semi-analytical formulation to show that the evolution of SIGOs generally proceeds to one of two states: either gravitational collapse to form stars outside of halos (followed by accretion onto halos as a cluster), or fall back into a nearby DM object, where the SIGO can evolve as substructure within the halo. This can be expressed as a timescale problem: SIGOs with a shorter collapse timescale and cooling timescale than fall-back timescale (to the nearest DM object) should form stars outside of halos.
Otherwise, we suggest that the SIGO can form a GC-like object, since GCs are often found near the edges of DM halos. Star formation may take place in this overdense gas at the outskirts of the halo. An example evolutionary path of a SIGO that has fallen back into a nearby halo is shown in Figure~\ref{Fig:SIGOFallback}. This SIGO had a low mass and an unusually low fall-back timescale (owing to a massive nearby halo), and therefore was unable to collapse outside of a DM halo. We also show, however, that a low initial mass may not prevent a SIGO from reaching the masses needed from star formation outside of halos. In Figure~\ref{Fig:SIGOGrowth}, we show an example of a SIGO that grows in mass by a factor of nearly $10$ over the course of the simulation, lowering its cooling time by $2$ orders of magnitude. This presents the possibility that even SIGOs that form outside of halos at lower masses may be capable of cooling to form stars outside of halos through growth, and suggests the need in future work for an analytic model for understanding SIGO growth rates.

Future studies aiming to investigate the collapse of SIGOs should carefully consider the cooling timescale (Figure~\ref{Fig:CollapseTimescale}) in determining a final redshift and resolution for their simulations. While individual SIGOs with unusually short cooling timescales may collapse prior to $z=20$, a typical early SIGO forming at $z=25$ with a typical cooling timescale of $100$ Myr will not fully collapse until around $z=17$ (or potentially later with lower values of $\sigma_8$). For similar reasons, it is important to account for the two-body relaxation timescale in simulations beyond $z=20$, which may be comparable to $100$ Myr at similar resolutions to those in this paper.

Explicit treatments of star formation with metal line cooling are needed to study the further evolution of SIGOs, including the size of their stellar populations. Exciting observational predictions may become possible with this next step, such as determining a half-light radius for SIGOs, and comparing it not only to present day GCs, but also to expected higher-redshift GC observations and potentially GC progenitor observations from JWST. These simulations will also begin to establish the efficiency of star formation in SIGOs, allowing us to understand characteristic maximum stellar population masses from the process. Stellar feedback, which is an important aspect of the later evolution of SIGOs, must also be considered, though it is beyond the scope of the current work. As these simulations develop, zoom-in simulations including radiative feedback should also be run to the present day, in order to give us a picture of the entire evolutionary history of a SIGO, allowing us to much more firmly connect these high-redshift objects to the structures we see in today's Universe.

\acknowledgments

The authors thank the anonymous reviewer for their helpful comments and suggestions. W.L., S.N., Y.S.C, B.B., F.M., and M.V. thank the support of NASA grant No. 80NSSC20K0500 and the XSEDE AST180056 allocation, as well as the
Simons Foundation Center for Computational Astrophysics and the UCLA cluster \textit{Hoffman2} for computational resources. F.M. acknowledges support from the Program ``Rita Levi Montalcini'' of the Italian MUR. S.N thanks Howard and Astrid Preston for their generous support. Y.S.C thanks the partial support from UCLA dissertation year fellowship. B.B. also thanks the the Alfred P. Sloan Foundation and the Packard Foundation for support. MV acknowledges support through NASA ATP grants 16-ATP16-0167, 19-ATP19-0019, 19-ATP19-0020, 19-ATP19-0167, and NSF grants AST-1814053, AST-1814259,  AST-1909831 and AST-2007355. NY acknowledges financial support from JST AIP Acceleration Research JP20317829.

\appendix
\section{Comparison of Important Timescales}\label{Sec:appendix}

\begin{figure*}[t]

\centering

\gridline{\fig{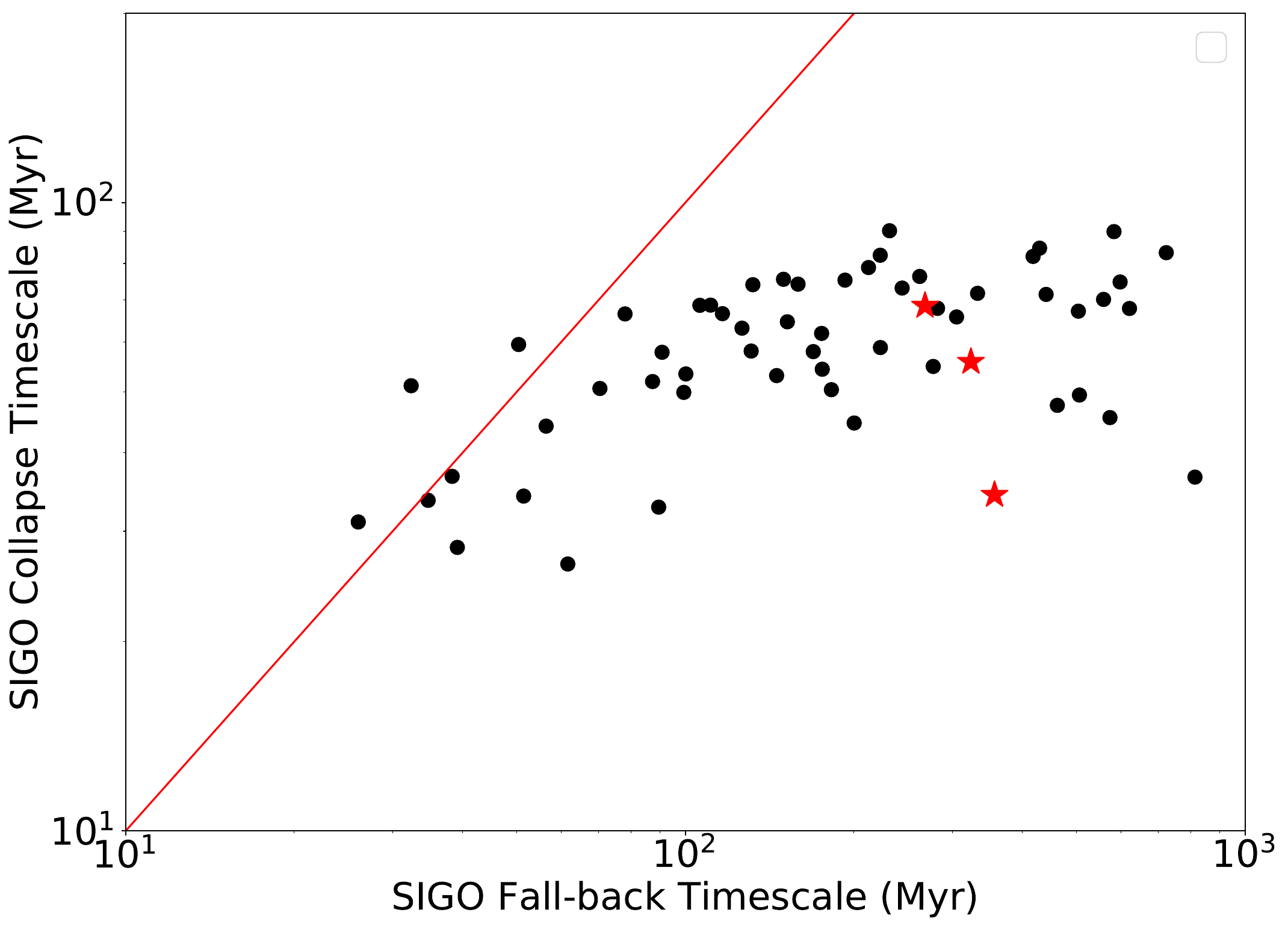}{0.47\textwidth}{\centering Collapse timescale vs. median fall-back timescale}
          \fig{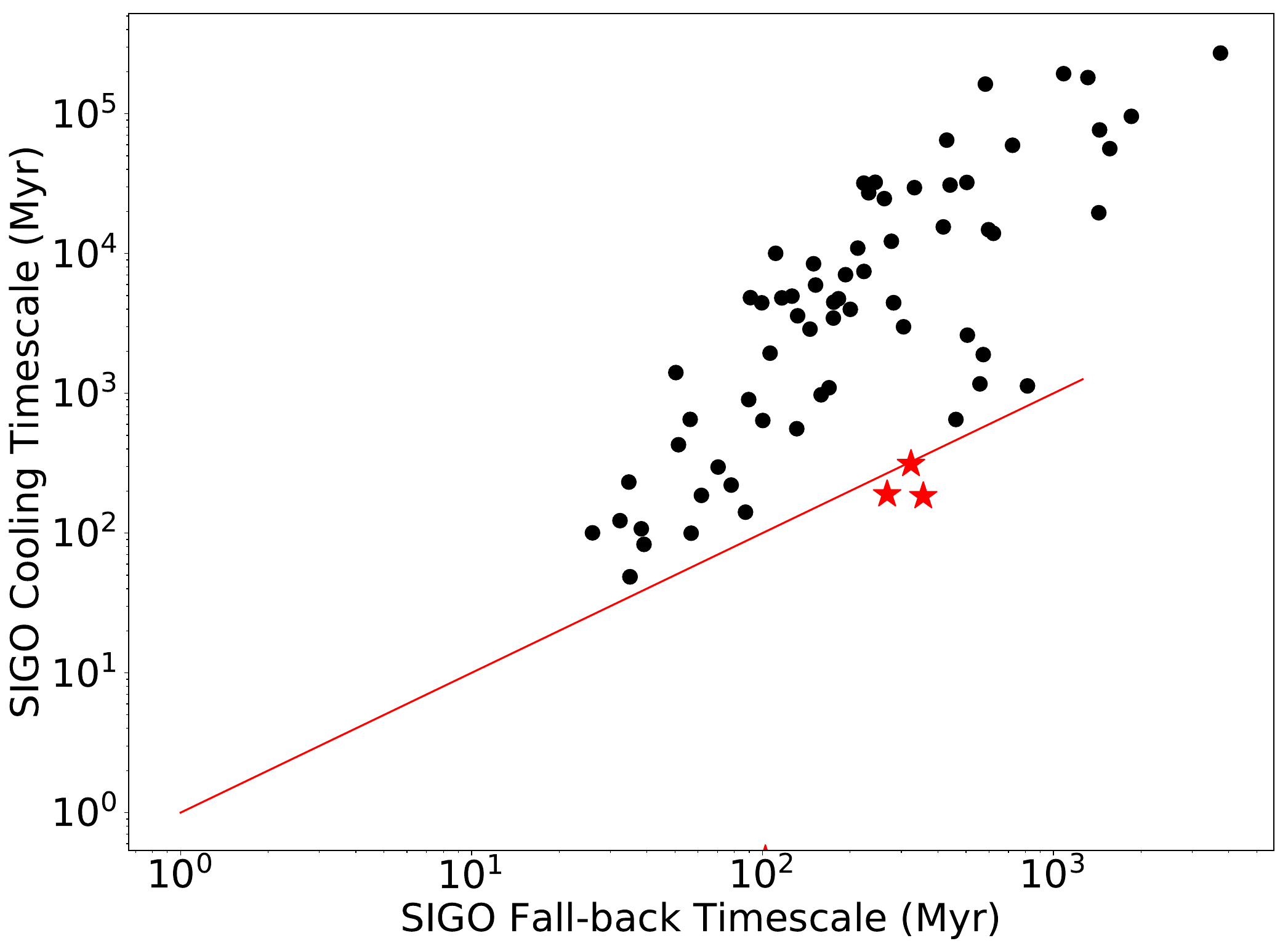}{0.47\textwidth}{\centering Cooling timescale vs. median fall-back timescale}
          }
          
\gridline{\fig{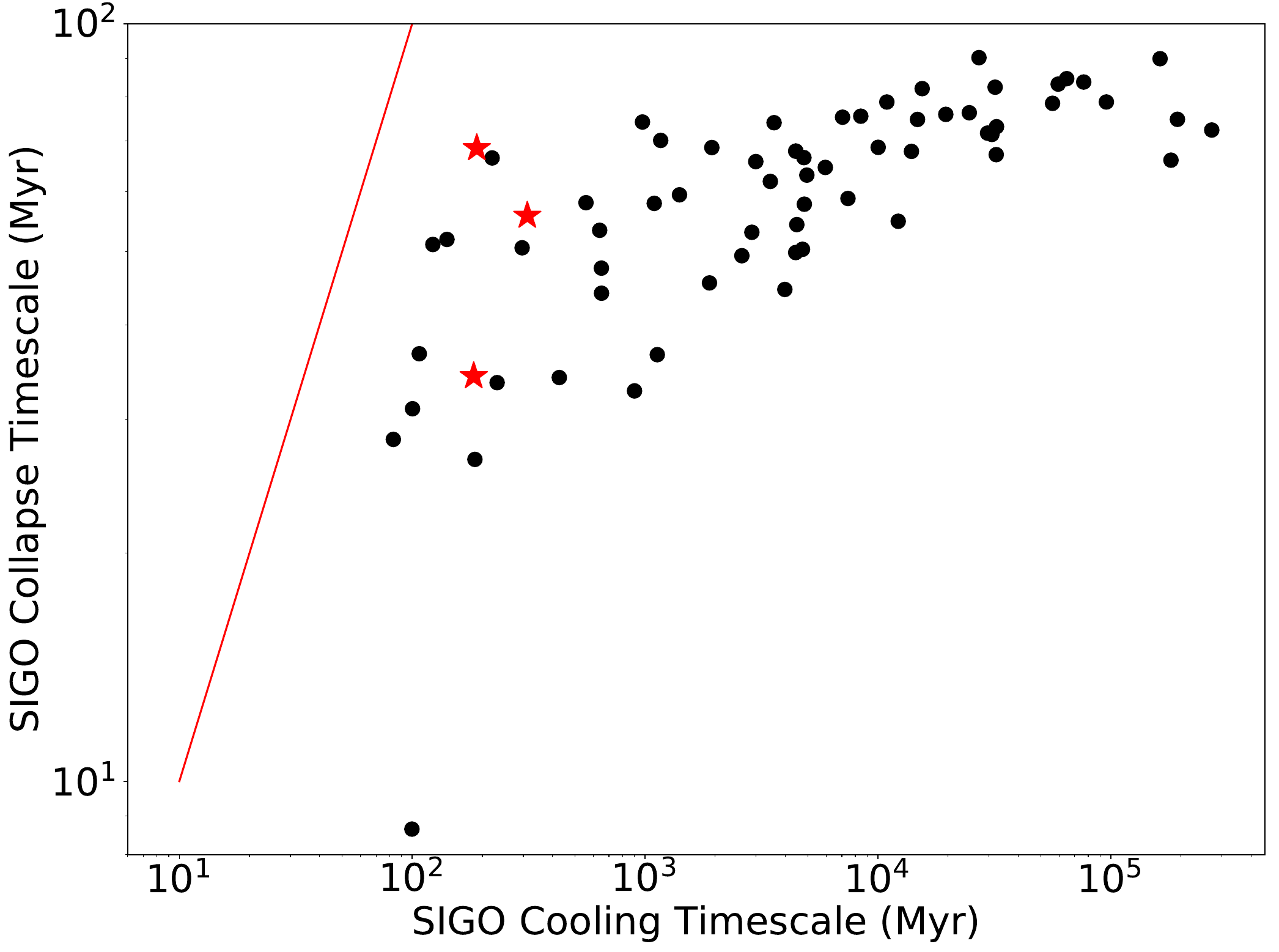}{0.47\textwidth}{\centering Collapse timescale vs. Cooling timescale}
          \fig{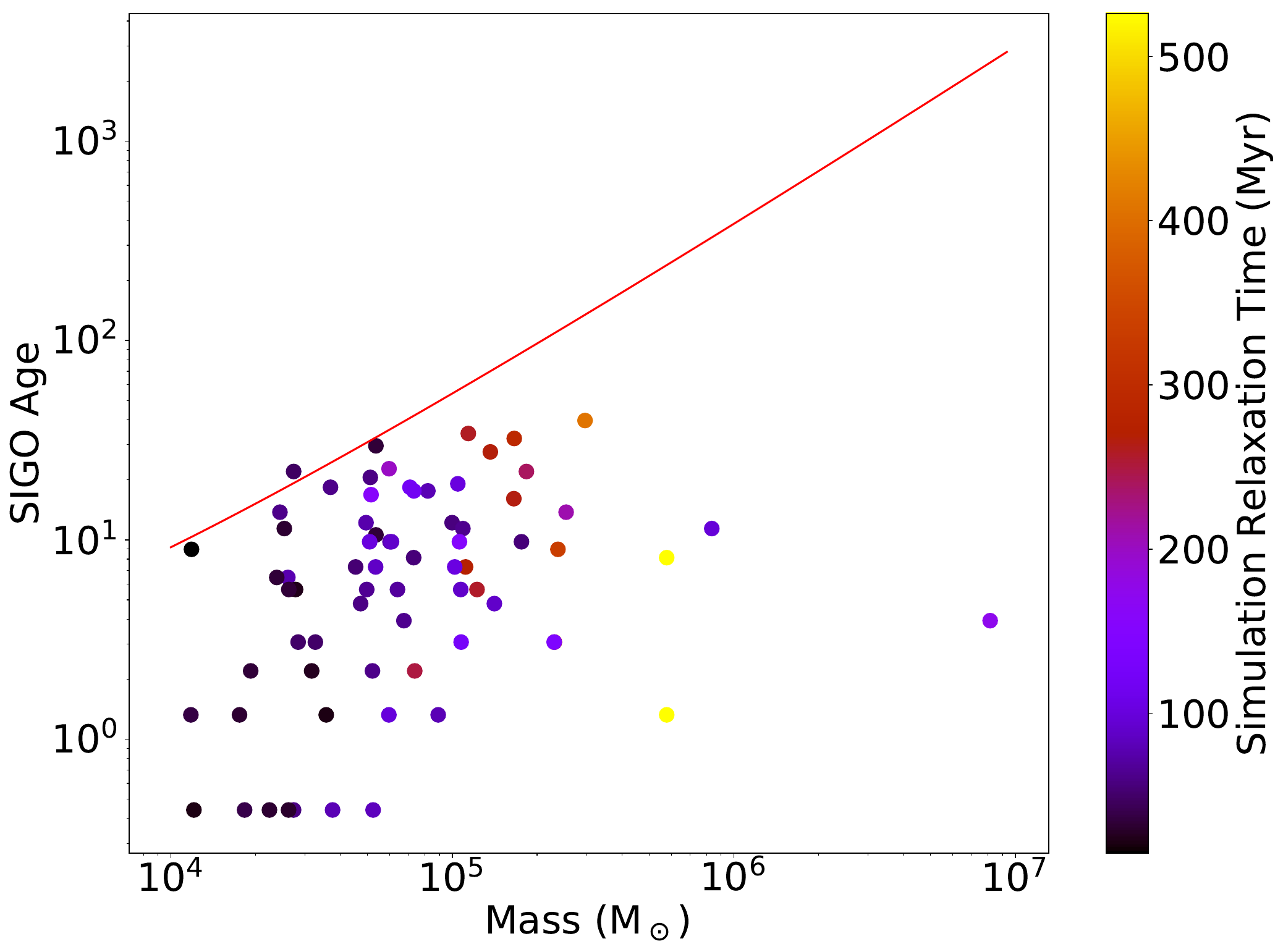}{0.47\textwidth}{\centering Relaxation timescale at simulation resolution}
          }
    \caption{{\bf Timescale Comparisons:} here we show a number of important timescales to the evolution of SIGOs at $z=20$ in Run $2$vH$2$. In the first $3$ panels (a-c), a black dot indicates a SIGO from the simulation. A red star indicates a SIGO that is likely to form stars outside of a DM halo, with cooling and collapse timescales shorter than its median fallback timescale. Note that $2$ star-forming SIGOs could not be pictured in this figure, due to their very short cooling and collapse timescales indicative that collapse has already occurred. The red lines in these figures indicate equality between the two timescales being compared. The final panel (d) shows the ages of $z=20$ SIGOs at the resolution of the simulation plotted against their masses. The red line shows the relaxation timescale at simulation resolution of a SIGO with typical properties as a function of mass.}\label{Fig:CollapseTimescale}
\end{figure*}

Figure~\ref{Fig:CollapseTimescale} displays another comparison of all of these timescales, as a proof of concept to aid the reader in gaining a physical intuition of how the various timescales compare. This allows a qualitative prediction of which SIGOs are able to collapse. The top left panel of the Figure shows the collapse timescale of SIGOs in Run $2$vH$2$. This is plotted against the SIGOs' median fall-back timescale (across time) to their nearest halo. This median corrects for errors in the FOF finder for DM halos: if a nearby halo merges with another, creating an ellipsoidal DM density distribution, the spherical treatment of halos in the simulation creates a deceptively long free-fall timescale, so it is preferable for the purposes of physical intuition of this evolution to refer instead to a median, which is more stable and not subject to the variations in halo-SIGO separation caused by mergers. As one can see, most SIGOs have a shorter collapse timescale than median fall-back timescale (and are below the red line in the panel reflecting equal timescales). This indicates that the cooling timescale is key to determining whether or not SIGOs will collapse before falling into nearby DM halos.

Panels (b) and (c) of Figure~\ref{Fig:CollapseTimescale} show two other comparisons of the timescales involved in this evolution. Panel (b) shows a comparison of the cooling timescale and the median fall-back timescale for all $z=20$ SIGOs. This panel further demonstrates that while the collapse timescale is important for the evolution of SIGOs, that timescale alone is usually insufficient to determine whether or not a SIGO will collapse independently of a DM halo-- the cooling timescale tends to be a bigger driver of such SIGOs' evolution, especially at lower masses. Panel (c) of Figure~\ref{Fig:CollapseTimescale} compares the cooling timescale for SIGOs to their collapse timescale, showing that the cooling timescale for SIGOs is, in fact, always longer than the collapse timescale at formation. Note that this cooling timescale includes an isochoric cooling term for the SIGO to begin to collapse (Eq. \ref{Eq:t_iso}): this does not reflect the balance between cooling times and collapse times once SIGOs achieve the Jeans mass and begin to collapse.

Unlike the other panels, Panel (d) of Figure~\ref{Fig:CollapseTimescale} shows a comparison of the relaxation timescale (Eq.~\ref{Eq:relaxtime}) of the simulation with mass, showing that we have lost some SIGOs at low masses to relaxation before $z=20$ in the simulation.

\begin{figure*}[t]

\centering

\gridline{\fig{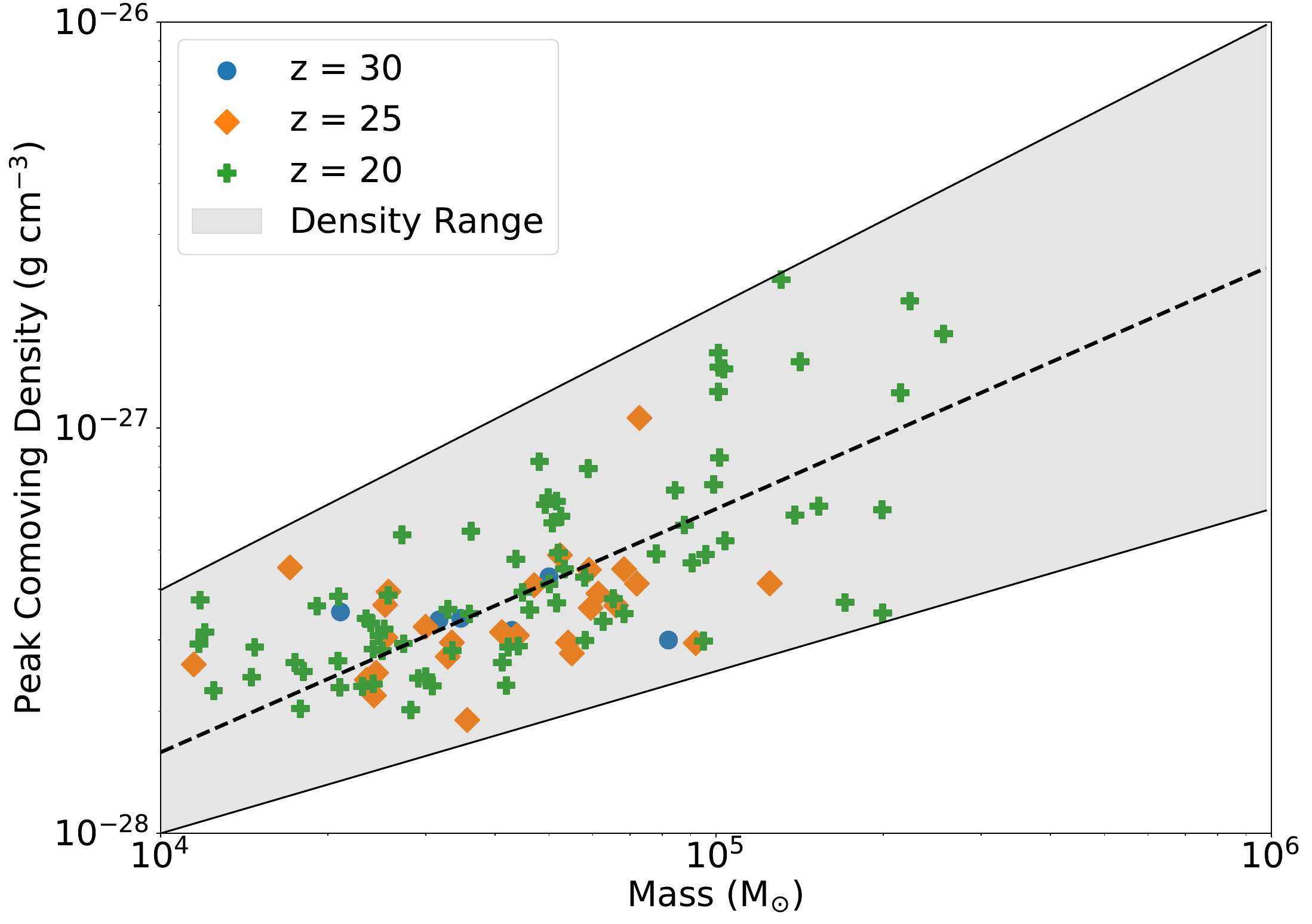}{0.47\textwidth}{}
          \fig{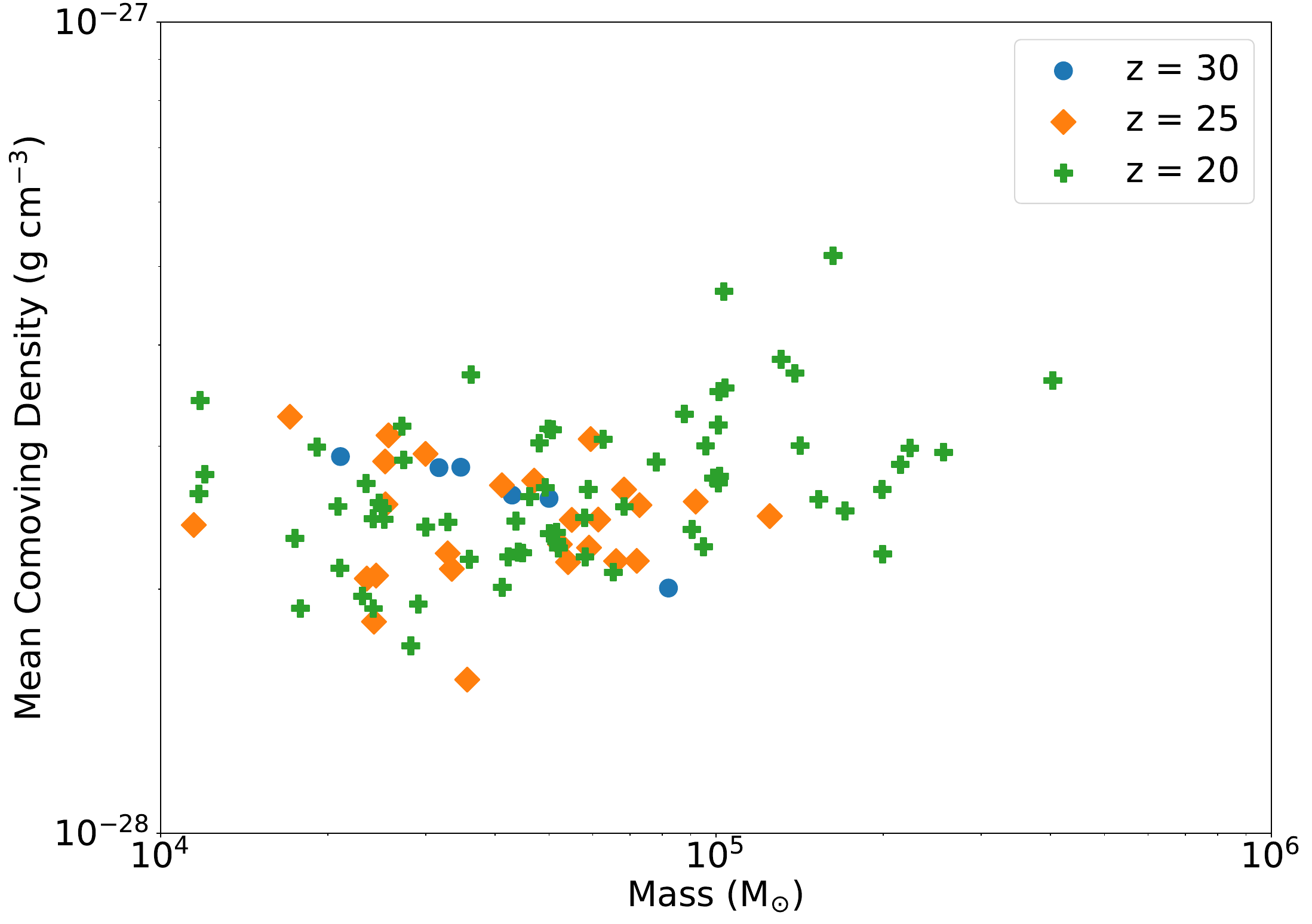}{0.47\textwidth}{}
          }
    \caption{{\bf Typical SIGO Densities:} here we show the peak (left panel) and mean (right panel) comoving densities of SIGOs at various redshifts in Run $2$vH$2$. In each panel, a mark indicates a SIGO from the simulation. The peak density is calculated as the volume-weighted average density of the most dense Voronoi gas cell in the SIGO and its $9$ nearest neighbours. The representative range of peak densities we use in this paper (Eqs. \ref{Eq:MaxDensity} and \ref{Eq:MinDensity}) is shown in the left panel in gray shading. The linear regression to the $z=20$ density data is presented as a dashed line. }\label{Fig:Densities}
\end{figure*}

SIGOs generally form in a similar range of densities regardless of mass. Their early evolution occurs primarily through cooling, and sometimes mass growth through gas accretion, rather than through immediate global collapse. However, SIGOs' collapse timescales depend on their mass, as their peak densities in their cores have a mass dependence. Figure~\ref{Fig:Densities} shows the relation between the peak and mean densities in SIGOs as a function of mass at several redshifts, to help illuminate this effect. In the right panel of this Figure, even SIGOs that have evolved for some time at $z=20$ do not have mean gas densities that exhibit a strong dependence on mass. This is expected: most SIGOs at this time are around $10$ Myr old, and nearly all of these SIGOs have not yet had a chance to globally cool and collapse on $\gtrapprox 50$ Myr timescales. However, there are hints of substructure formation in high-mass SIGOs. In the left panel of Figure~\ref{Fig:Densities}, we show the trend in peak density with mass, excepting $4$ $z=20$ SIGOs that have already partially or fully collapsed and reached much higher densities. A linear regression yields an approximate formula for the relation between peak comoving density and mass at $z=20$: log$_{10}$($\rho_{\rm peak,com}$ (g/cm$^3$)) = $0.6\times$log$_{10}$(M$_{\rm SIGO}$(M$_\odot$)) - $30.2$. As one can see, there is a clear positive correlation between peak density and mass--one that becomes stronger at later redshifts. At $z=30$, the Spearman coefficient is $-0.37$ ($N=6$, $p=0.47$), at $z=25$ it is $0.50$ ($N=28$, $p=0.007$), and at $z=20$ it is $0.78$ ($N=85$, $p=4\times10^{-19}$). Notably, this trend remains even if the highest-mass SIGOs at $z=20$ are excluded, indicating that it is not only a function of increased structure formation. If only SIGOs with a mass lower than the highest mass $z=25$ SIGO are studied, the Spearman coefficient at $z=20$ is $0.70$ ($N=69$, $p=1.7\times10^{-11}$). 

\begin{wrapfigure}{r}{8.cm}\setlength{\abovecaptionskip}{-15pt}\setlength{\belowcaptionskip}{30pt}
\centering
\gridline{\fig{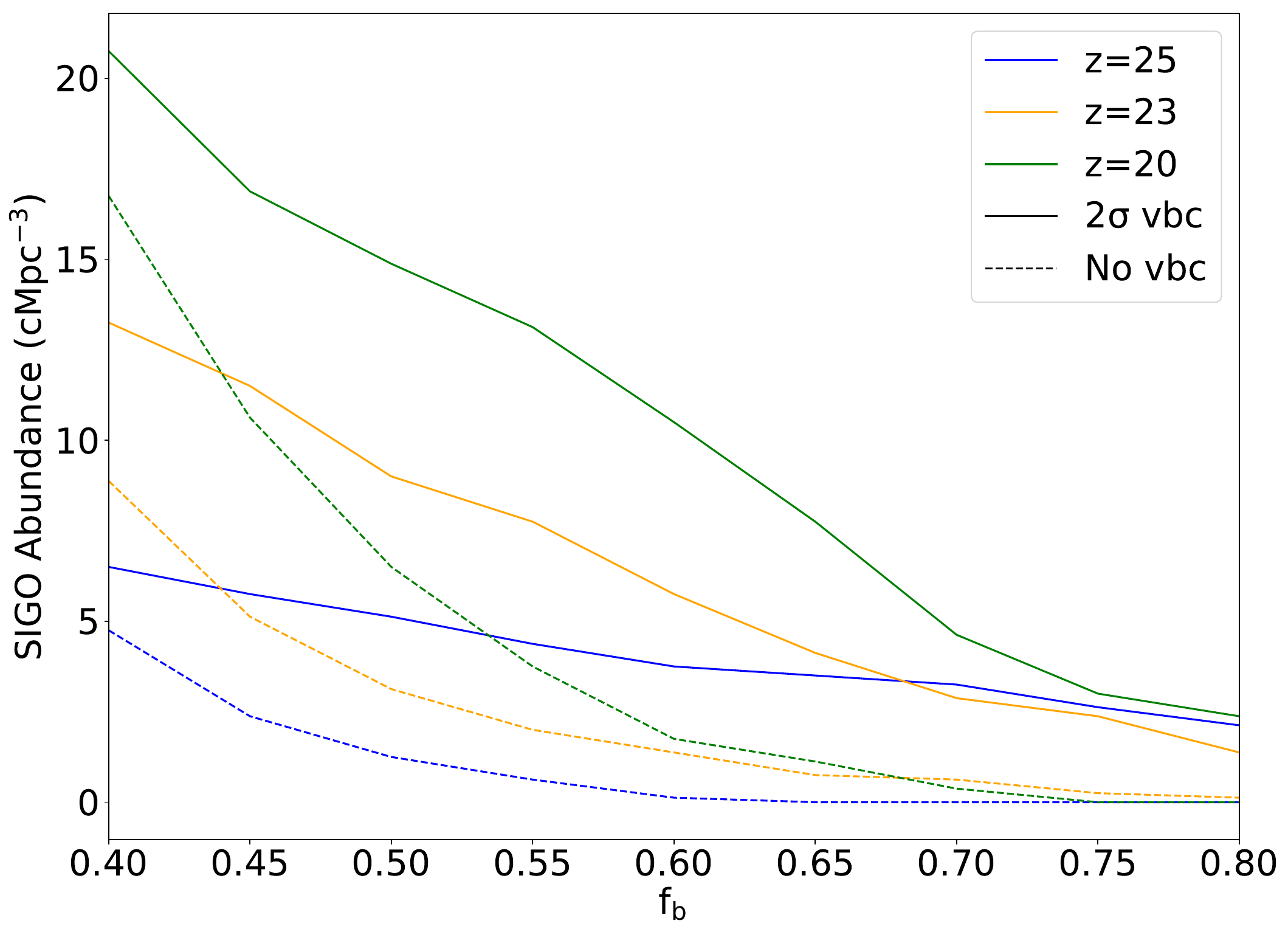}{.44 \textwidth}{}}
    \caption{{\bf Choosing a baryon fraction cutoff:} in this Figure we show the effect of different baryon fraction cutoffs on calculated SIGO abundances at $z=25$ (blue), $z=23$ (orange), and $z=20$ (green). Results from Run $2$vH$2$ are shown with solid lines, whereas results from Run $0$vH$2$ are shown with dashed lines. This Figure aims to justify our choice in f$_{\rm b}$ cutoff. } \label{Fig:VaryingFb}
\end{wrapfigure}

It is also important to consider the effects of varying the baryon fraction cutoff on SIGO abundances. In Figure~\ref{Fig:VaryingFb}, we show the effect of varying this cutoff with and without v$_{\rm bc}$, at a variety of redshifts when SIGOs from our $z=20$ snapshot formed. Below f$_{\rm b}$ = $0.6$, the abundances of SIGOs removed by increasing our f$_{\rm b}$ cutoff are comparable in our v$_{\rm bc}$ = $0$ and 2$\sigma$ runs, indicating that raising this cutoff helps us to distinguish spurious objects from true SIGOs. However, above f$_{\rm b}$ = 0.6, many more SIGOs are being removed in the 2$\sigma$ simulation than spurious objects from the $0$ v$_{\rm bc}$ simulations, indicating that we do not benefit from further raising the f$_{\rm b}$ cutoff at these redshifts. Notably, at $z=25$, this threshold is sufficient to remove all but 1 spurious SIGO. There are still a small fraction of spurious objects within the SIGO population at $z=20$, however, so the SIGOs which are marked as “potential star-forming SIGOs” in Figure~\ref{Fig:AgeVsMass} were then verified by ensuring that no ``SIGOs” in the no stream velocity simulation existed nearby (in the vicinity of the same identified parent halo).

\bigskip
\bigskip

\bibliography{myBib}{}
\bibliographystyle{aasjournal}



\end{document}